\documentclass[floatfix,twocolumn,amsmath,amssymb,reprint,footinbib,superscriptaddress,showpacs,longbibliography]{revtex4-1}
\usepackage{graphics}      
\usepackage{graphicx}      
\usepackage{longtable}     
\usepackage{url}           
\usepackage{bm}    
\usepackage{braket}        
\usepackage{xcolor}
\usepackage{float}  
\usepackage{natbib}
\usepackage{graphicx}
\usepackage{comment}
\usepackage{siunitx}
\usepackage{appendix}
\usepackage{amsmath,amssymb,epsfig,color}

\usepackage{soul}
\usepackage{cancel}

\usepackage{ulem} 

\usepackage{xcolor}
\usepackage{framed}
\definecolor{shadecolor}{rgb}{0.9,0.9,0.9}

\newcommand{\nn}{\nonumber}
\newcommand{\nl}{\nonumber \\}
\newcommand{\be}{\begin{equation}}
\newcommand{\ee}{\end{equation}}
\newcommand{\bea}{\begin{eqnarray}}
\newcommand{\eea}{\end{eqnarray}}

\begin{document}

\title {Perturbative Framework for Engineering Arbitrary Floquet Hamiltonian}

\author{Yingdan Xu}
\affiliation{Center for Joint Quantum Studies and Department of Physics, School of Science, Tianjin University, Tianjin 300072, China}

\author{Lingzhen Guo}
\thanks{lingzhen\_guo@tju.edu.cn}
\affiliation{Center for Joint Quantum Studies and Department of Physics, School of Science, Tianjin University, Tianjin 300072, China}

\begin{abstract}
We develop a systematic perturbative framework to engineer an arbitrary target Hamiltonian in the Floquet phase space of a periodically driven oscillator based on Floquet-Magnus expansion. The high-order errors in the engineered Floquet Hamiltonian are mitigated by adding high-order driving potentials perturbatively. {We introduce a transformation method that allows us to obtain an analytical expression of the leading-order correction drive for engineering a target Hamiltonian with discrete rotational and chiral symmetries in phase space. We also provide a numerically efficient procedure to calculate high-order correction drives and apply it to engineer the target Hamiltonian with degenerate eigenstates of multi-component cat states that are important for fault-tolerant hardware-efficiency bosonic quantum computation.} 
\end{abstract}

\date{\today}

\maketitle

\section{Introduction}\label{sec-intro}
Floquet systems with periodic drives provide versatile platforms to investigate novel physics that are not accessible for static systems. A range of intriguing physical phenomena, such as Floquet topological physics \cite{jiang2011prl,rudner2013prx,hu2015prx} and Floquet/discrete time crystals \cite{sacha2015pra,else2016prl,yao2017prl,Zhang2017nat,Sacha2018rpp}, have attracted extensive attentions in recent years. Meanwhile, the periodic drive is a ubiquitous recipe to engineer quantum systems for quantum technologies \cite{marin2015aip,eckardt2017rmp}. Therefore, it is  of importance not only in theory but also in practice to understand the role of periodic drive.

Floquet theory claims that the stroboscopic dynamics of a periodically driven system can be described effectively by a time-\textit{independent}  Floquet Hamiltonian \cite{Floquet1883,Shirley1965pr}. However, it is in general impossible to obtain the exact analytical form of Floquet Hamiltonian except for very few simple models. Fortunately, the Magnus theorem provides a perturbative tool for calculating the Floquet Hamiltonian in a series of inverse driving frequencies \cite{Blanes2009PR}. Other well-known perturbative frameworks to calculate the effective Hamiltonian (up to a gauge difference from the Floquet Hamiltonian) include the van Vleck degenerate perturbation theory \cite{Eckardt2015NJP} and the Brillouin-Wigner perturbation theory \cite{Mikami2016prb}.

Floquet engineering, which aims to design a proper driving scheme such that the corresponding Floquet or effective Hamiltonian approaches the desired target Hamiltonian, is a very developed and active research field \cite{marin2015aip,Rudner2020nrp,Jangjan2020scirep,zemlevskiy2024pra}. 
An important application of Floquet engineering is to generate nonclassical bosonic states \cite{Gerry2004book,Strekalov2019springer,Kubala2015njp} with discrete translational or rotational symmetries in phase space \cite{Leghtas2013prl,Heeres2017nc,Rosenblum2018science,Fluhmann2019nature,Hu2019nature,Campagne-Ibarcq2020nature,Gertler2021nature},  
 for hardware-efficiency quantum error correction \cite{Tzitrin2020pra,Terhal2020iop,Joshi2021qst,weizhou2021fr} and fault-tolerant bosonic quantum computation \cite{Cochrane1999pra,Gottesman2001pra,Travaglione2002pra,Michael2016prx,Albert2018pra,Arne2020prx}. By designing a proper driving protocol  \cite{Puri2019PRX,Rymarz2021prx,Conrad2021pra,xanda2023arxiv}, specific target bosonic code states can be prepared and stabilized against various noises in the environment. 
 For instance, one can prepare the Gottesman-Kitaev-Preskill (GKP) state via dynamical decoupling \cite{Conrad2021pra} or adiabatic ramp \cite{xanda2023arxiv}, and stabilize the Schr\"odinger-cat state against phase-flip errors \cite{Puri2019PRX}. 
In a recent paper \cite{guo2024prl}, we proposed a general method of \textit{arbitrary phase-space Hamiltonian engineering} (APSHE) that can engineer arbitrary Hamiltonians in the Floquet phase space of a periodically driven oscillator. Combined with the adiabatic ramp protocol \cite{xanda2023arxiv}, our APSHE method can be exploited to prepare arbitrary desired quantum bosonic code state. 

However, most of the works so far have focused on the implementation of specific target Hamiltonians or bosonic code states of interest with the rotating wave approximation (RWA) or lowest-order Floquet-Magnus expansions. The ignored high-order non-RWA terms cause errors in Floquet engineering. In order to cancel the errors beyond RWA, additional correction driving terms are needed. However,  the additionally added driving terms usually introduce more errors to the engineered Hamiltonian or states.  
\textit{Does there exist a systematic method to construct high-order drives that can mitigate the errors from higher-order Floquet-Magnus expansions up to desired precision?}   This is the inverse Floquet-Magnus problem for designing arbitrary Floquet Hamiltonian with arbitrary precision. 

In this work, we provide such a method for a single quantum particle by developing a systematic perturbative framework to calculate the drives that can approach a given arbitrary Floquet Hamiltonian up to desired-order precision. Especially,  we introduce a transformation that can circumvent the difficulty of calculating high-order commutators in the higher-order Floquet-Magnus expansions and directly construct the additional correction driving terms. 
{
We apply our method to the model of a monochromatically driven oscillator and obtain an analytical expression for the leading-order correction, which is then verified by numerical results.  We also provide a numerically efficient procedure to engineer a target Hamiltonian with degenerate eigenstates of multi-component cat states. Our method offers a powerful tool for generating nonclassical quantum states for fault-tolerant bosonic quantum computation in a range of experimental platforms such as superconducting circuits with Josephson junctions.

}

%
\section{General theory}\label{sec-general}
\subsection{Model and Goal}\label{sec-mg}
We consider a periodically driven  oscillator with the Hamiltonian described by 
\bea\label{eq-HtVx}
\mathcal{H}(t)=\frac{1}{2}(\hat{x}^2+\hat{p}^2)+ V(\hat{x},t).
\eea
Here, all the variables have been scaled dimensionless by the characteristic units of the system. The position and momentum are scaled such that $[\hat{x},\hat{p}]=i\lambda$, where $\lambda$ is the dimensionless Planck constant. 
We define in passing the ladder operator $\hat{a}\equiv(\hat{x}+i\hat{p})/\sqrt{2\lambda}$ with $[\hat{a},\hat{a}^\dagger]=1$.
The units of energy (Hamiltonian), frequency, and time are set to be $\hbar\omega_0\lambda^{-1}$, $\omega_0$ and $\omega^{-1}_0$ respectively, where $\omega_0$ is the harmonic oscillator frequency.  In our model, the nonlinearity of the oscillator is incorporated in the potential $V(\hat{x},t)$ that can include static terms.

We assume the driving field has frequency $\omega_d$, i.e., $V(\hat{x},t)=V(\hat{x},t+T_d)$ with $T_d=2\pi/\omega_d$ defined as the period of driving field. 
To proceed, we work on the multi-photon resonance condition that the driving frequency is set to be $n$ times the natural frequency of the harmonic oscillator, i.e., $T_d=2\pi/n$ with $n\in \mathbb{Z}^+$. Note that any integer multiple period $T=nT_d $ $(n\in \mathbb{Z}^+)$ is also the driving period, i.e., $V(\hat{x},t)=V(\hat{x},t+T)$.
By transforming into the rotating frame with frequency $\Omega=2\pi/T$, we have 
$\hat{O}(t)\hat{x}\hat{O}^\dagger(t)=\hat{x}\cos (\Omega t)+\hat{p}\sin (\Omega t)$  with free time-evolution operator $\hat{O}(t)\equiv e^{i\hat{a}^\dagger\hat{a}\Omega t}$ and the Hamiltonian in the rotating frame
\bea\label{eq-Ht}
\hat{H}(t)&\equiv&\hat{O}(t)\mathcal{H}(t)\hat{O}^\dagger(t)-i\lambda O(t)\dot{O}^\dagger(t)\nl
&=& V\Big[\hat{x}\cos (\Omega t)+\hat{p}\sin (\Omega t),t\Big]\nl
&\equiv&
\sum_{l\in \mathbb{Z}}\hat{H}_le^{il\Omega t}.
\eea
Here, $\hat{H}_l(\hat{x},\hat{p})=T^{-1}\int_0^T \hat{H}(t)e^{-il\Omega t}dt$ is the decomposed harmonics of the rotating-frame Hamiltonian. According to the hermiticity of  $\hat{H}(t)$, we have the important relationship
$\hat{H}_l^\dagger(\hat{x},\hat{p})=\hat{H}_{-l}(\hat{x},\hat{p})$  for the Hamiltonian harmonics.

Floquet theory claims that the stroboscopic time evolution of a periodically driven system is described by a time-independent \textit{Floquet Hamiltonian} $\hat{H}_F$ determined by \cite{Liang2018njp}
\bea\label{eq-HFt0}
e^{-i\frac{T}{\lambda}\hat{H}_F(t_0)}\equiv U(t_0+T,t_0)=\mathcal{T}e^{-i\frac{1}{\lambda}\int_{t_0}^{t_0+T}\hat{H}(t)dt}, 
\eea
where $\mathcal{T}$ is the time-ordering operator. The Floquet Hamiltonian $\hat{H}_F(t_0)$ describes the stroboscopic time evolution starting from the initial reference time $t_0$ with stroboscopic time step $T$. Note that the eigenstates of Floquet Hamiltonian $\hat{H}_F(t_0)$ depend on the choice of the initial reference time $t_0$. However, according to  the Floquet theorem \cite{Floquet1883,Shirley1965pr,Sambe1973pra,Grifoni1998pr,Eckardt2015NJP}, the eigenvalues of $\hat{H}_F(t_0)$ should be free of the choice of reference time $t_0$. We will elucidate this subtle point later in Section~\ref{sec-heff}.

In general, Floquet Hamiltonian $\hat{H}_F(\hat{x},\hat{p})$ for a fixed initial time $t_0$ is an arbitrary function of noncommutative operators $\hat{x}$ and $\hat{p}$,  that cannot be simply decomposed into the sum of kinetic and potential terms. Except for very few models, it is impossible to obtain an exact form of Floquet Hamiltonian.
Fortunately, in the regime where the driving frequency $\omega_d$ (and the chosen Floquet frequency $\Omega$) is much larger than the characteristic frequency of the system, the Floquet Hamiltonian can be given in the so-called Floquet-Magnus expansion \cite{Casas2001NJP,Blanes2009PR} $\hat{H}_F=\sum_{n=0}^\infty \hat{H}^{(n)}_F$ in the order of perturbative parameter $\Omega^{-1}$. The leading-order term $\hat{H}^{(0)}_F(\hat{x},\hat{p})$ is just the time averaged Hamiltonian $\hat{H}(t)$ over one Floquet period $T$ 
\bea\label{eq-h0h1h3}
\hat{H}^{(0)}_F(\hat{x},\hat{p})&=&\frac{1}{T}\int_{t_0}^{t_0+T}dt\hat{H}(t)=\hat{H}_0.
\eea
This is also the effective Hamiltonian in the RWA obtained from all the other perturbative methods \cite{Eckardt2015NJP,Mikami2016prb}. Note that the RWA Floquet Hamiltonian $\hat{H}_0$ is independent of initial time choice $t_0$, cf. Eq.~(\ref{eq-Ht}). Higher-order Floquet-Magnus expansion terms can be expressed with the periodic Hamiltonian harmonics $\hat{H}_l$ \cite{Casas2001NJP,Blanes2009PR,Eckardt2015NJP,Mikami2016prb}, cf. Eq.~(\ref{eq-H1Fxp}) below for the first-order Magnus expansion.

The goal of this work is to engineer the real-space driving potential $V(\hat{x},t)$ to generate an arbitrary target Hamiltonian $H_T(\hat{x},\hat{p})$ in phase space beyond RWA. We provide a general perturbative procedure for the calculation of driving potential $V(\hat{x},t)=\sum_{i=0}V^{(i)}(\hat{x},t)$ that can mitigate high-order Floquet-Magnus expansions and make the Floquet Hamiltonian $\hat{H}_F(\hat{x},\hat{p})$ approaching the target Hamiltonian $\hat{H}_T(\hat{x},\hat{p})$ up to desired order of perturbative parameter $\Omega^{-1}$.  

\subsection{Noncommutative Fourier transformation }\label{sec-ncft}
 For a given target Hamiltonian  $\hat{H}_T(\hat{x},\hat{p})$, we introduce a Fourier decomposition of target Hamiltonian by writing it as a sum of plane-wave operators in the noncommutative phase space  \cite{guo2024prl}, i.e.,
\bea\label{eq-HTxp}
\hat{H}_T(\hat{x},\hat{p})
=
\frac{1}{2\pi}\int \int dk_x dk_pf_T(k_x,k_p)e^{i(k_x\hat{x}+k_p\hat{p})}
\eea
with $f_T(k_x,k_p)$  the \textit{noncommutative Fourier transformation} (NcFT) coefficient.
In order to calculate Eq.~(\ref{eq-HTxp}) analytically, we write the target Hamiltonian with reordered ladder operators as 
$$\hat{H}_T(\hat{a}^\dagger,\hat{a})\equiv\sum_{n,m}\chi_{nm} (\hat{a}^{\dagger})^n\hat{a}^m.$$
Note that the \textit{ordering} here keeps all the terms from commutators (e.g., $\hat{a}\hat{a}^\dagger=\hat{a}^\dagger\hat{a}+1$) and is different from the \textit{normal ordering} ($:\hat{a}\hat{a}^\dagger:=\hat{a}^\dagger\hat{a}$) in the study of quantum field theory \cite{Greiner1996}. 

Using the coherent state $|\alpha\rangle$ defined as the eigenstate of lowering operator via $\hat{a}|\alpha\rangle=\alpha |\alpha\rangle$, we calculate the Q-function of the Hamiltonian operator in the coherent representation as follows
$$H_T(\alpha,\alpha^*)=\langle \alpha |\hat{H}_T|\alpha \rangle=\sum_{n,m}\chi_{nm}(\alpha^*)^n\alpha^m.$$
By identifying $\alpha=(x+ip)/\sqrt{2\lambda}$ with $x\equiv\langle \alpha |\hat{x}|\alpha \rangle$ and $p\equiv\langle \alpha |\hat{p}|\alpha \rangle$, we can write the Hamiltonian Q-function in phase space as $H_T(x,p)$. 
Then using the identity \cite{Liang2018njp,guo2024prl}
$$
\langle \alpha|e^{i(k_x\hat{x}+k_p\hat{p})}|\alpha\rangle=e^{-\frac{\lambda}{4}(k^2_x+k^2_p)}e^{i(k_xx+k_pp)},
$$
we obtain the NcFT coefficient in Eq.~(\ref{eq-HTxp}) as follows \cite{guo2024prl}
\bea\label{eq-fFT}
f_T(k_x,k_p)=\frac{e^{\frac{\lambda}{4}(k^2_x+k^2_p)}}{2\pi}\int\int dxdp H_T(x,p) e^{-i(k_xx+k_pp)}.\  
\eea
From the hermiticity of Hamiltonian operator $\hat{H}_T=\hat{H}^\dagger_T$, the NcFT coefficient satisfies $f(k_x,k_p)=f^*(-k_x,-k_p)$. 
The above NcFT technique differs from the conventional Fourier transformation on two points: (1) there is an additional factor $e^{\frac{\lambda}{4}(k^2_x+k^2_p)}$; (2) one has to reorder the ladder operators in the target Hamiltonian. Our NcFT technique can be viewed as a variant of quantum distribution theory \cite{scully1997quantum}.

 We can also transform into the polar coordinate system by introducing $(k_x=k\cos\tau, k_p=k\sin\tau)$, and rewrite the Fourier series expansion Eq.~(\ref{eq-HTxp}) as
\bea\label{eq-Fktheta}
\hat{H}_T&=&\frac{1}{2\pi}\int_{0}^{2\pi}d\tau\int_{-\infty}^{+\infty}dk\frac{|k|}{2}f_T(k,\tau)e^{ik(\hat{x}\cos\tau+\hat{p}\sin\tau)}.\ \ \ \ 
\eea
Here, we have defined the NcFT coefficient in the polar coordinate system via 
\bea\label{eq-FTktau}
f_T(k,\tau)\equiv f_T(k_x,k_p)
\eea
allowing for $k<0$ via the relation $f_T(k,\tau)\equiv f^*_T(-k,\tau)$. 
Having the NcFT coefficient $f_T(k,\tau)$ of target Hamiltonian $\hat{H}_T(\hat{x},\hat{p})$, we engineer the zeroth-order (with respect to the parameter  $\Omega^{-1}$) real-space driving potential $V(x,t)=V^{(0)}(x,t)$ in Eq.~(\ref{eq-HtVx})  as follows
\bea\label{eq-Vxt-1}
V^{(0)}(x,t)&=&\int_{-\infty}^{+\infty}\frac{|k|}{2}f_T(k,\Omega t)e^{i kx}dk.
\eea
Here, note that we ignore the hat of the position operator because there is no moment operator (we will keep this notation below for simplicity).
In the rotating frame with frequency $\Omega$, the corresponding rotating-frame Hamiltonian, cf. Eq~(\ref{eq-Ht}), becomes 
\bea\label{eq-Ht-ft}
H(t)&=&\int_{-\infty}^{+\infty}\frac{|k|}{2}f_T(k,\Omega t)e^{ik[\hat{p}\sin (\Omega t)+\hat{x}\cos (\Omega t)]}dk.\ \ \ 
\eea
From Eqs.~(\ref{eq-h0h1h3}), (\ref{eq-Fktheta}) and (\ref{eq-Ht-ft}), the lowest-order Magnus expansion of Floquet Hamiltonian is just the target Hamiltonian 
$H^{(0)}_F(\hat{x},\hat{p})=\hat{H}_T(\hat{x},\hat{p})$
by identifying the parameter $\tau=\Omega t$ \cite{guo2024prl}.

The driving potential given by Eq.~(\ref{eq-Vxt-1}) can be engineered by superposing a series of cosine lattice potentials 
\bea\label{eq-Vxt-2}
V^{(0)}(x,t)
&=&\int_{-\infty}^{+\infty}A(k, t)\cos[kx+\phi(k,t)]dk.
\eea
with tunable amplitudes $A(k,t)=|kf_T(k,\Omega t)|$ and phases $\phi(k,t)=\mathrm{Arg}[f_T(k,\Omega t)]$ depending on time and wave vector $k$. Such driving scheme can be implemented in the cold atom experiments with optical lattices that are formed by laser beams intersecting at an angle \cite{Moritz2003prl,Hadzibabic2004prl,Guo2022prb} or in the sperconducting circuits \cite{Chen2014prb,Hofheinz2011prl,Chen2011apl} with dc-voltage biased Josephson junctions  \cite{Armour2013prl,Gramich2013prl,Juha2013prl,Juha2015prl,Juha2016prb,Armour2015prb,Trif2015prb,Kubala2015iop,Hofer2016prb,Dambach2017njp,Lang2021njp,Lang2022arxiv}. 
%
%

\subsection{Perturbative framework}\label{sec-perfra}

We emphasize that the above Floquet Hamiltonian engineering method relies on the RWA, cf. Eq.~(\ref{eq-h0h1h3}), which is the lowest-order Floquet-Magnus expansion.
However, as the zeroth-order engineered driving potential $V^{(0)}(x,t)$ in general also contains high-order subharmonics, cf. Eq.~(\ref{eq-Ht}), the corresponding high-order Flqouet-Magnus expansions eventually deviate the exact Floquet Hamiltonian away from the target Hamiltonian $\hat{H}_T(\hat{x},\hat{p})$. To mitigate the higher-order Floquet-Magnus terms, we need to introduce additional correction driving potentials. 
First, we show how to cancel the 1st-order Floquet-Magnus expansion ($\propto \Omega^{-1}$) given by \cite{Mikami2016prb}
\bea\label{eq-H1Fxp}
\hat{H}^{(1)}_F(\hat{x},\hat{p})&=&\frac{1}{\lambda\Omega}\sum_{l\neq 0}^\infty\Big(\frac{1}{2l}[\hat{H}_l,\hat{H}_{-l}]+\frac{1}{l}[\hat{H}_{-l},\hat{H}_{0}]e^{il\Omega t_0}\Big)\nl
&=&\frac{1}{\lambda\Omega}\sum_{l=1}^\infty\Big(\frac{1}{2l}[\hat{H}_l,\hat{H}_{-l}]+\frac{1}{l}[\hat{H}_{-l},\hat{H}_{0}]e^{il\Omega t_0}\Big)\nl
&&+h.c.,
\eea
where we have used the property $\hat{H}_l^\dagger=\hat{H}_{-l}$, cf. the discussion below Eq.~(\ref{eq-Ht}). Note that $\hat{H}^{(1)}_F(\hat{x},\hat{p})$ depends on the initial reference time $t_0$. %
 In fact, if we shift the initial time (driving phase) of the periodically driven Hamiltonian by $\hat{H}(t)\rightarrow\hat{H}(t+t_0)$,  the harmonics of shifted Hamiltonian follow $\hat{H}_l\rightarrow\hat{H}_le^{-il\Omega t_0}$ according to Eq.~(\ref{eq-Ht}), and the $t_0$-dependent terms in the above 1st-order Floquet-Magnus Hamiltonian are canceled.

By calculating the NcFT coefficient $f^{(1)}(k,\Omega t)$ of Hamiltonian $ H^{(1)}_F(\hat{x},\hat{p})$, we introduce the additional 1st-order correction driving potential as follows
\bea\label{eq-V1xt}
V^{(1)}(x,t)=-\int_{-\infty}^{+\infty} \frac{|k|}{2}f^{(1)}(k,\Omega t)e^{i kx}dk.
\eea
Note that there is a minus sign in front compared to the zeroth-order driving potential given by Eq.~(\ref{eq-Vxt-1}).
According to our previous discussion, the RWA Floquet Hamiltonian from $V^{(1)}(x,t)$ will cancel the 1st-order Floquet-Magnus expansion given by Eq.~(\ref{eq-H1Fxp}).
 Now the total driving potential becomes
$$
V(x,t)=V^{(0)}(x,t)+V^{(1)}(x,t).
$$
In general, the additional driving field $V^{(1)}(t)$ also introduces high-order Floquet-Magnus expansion terms ($\propto \Omega^{-m}$ with $m\geq 2$) . 

To build the perturbative framework by constructing high-order driving potentials, we define the harmonics $\hat{V}^{(m)}_l$ of $m$-th order driving potential $V^{(m)}(x,t)$ in the rotating frame with frequency $\Omega$ by
\bea\label{eq-Vil}
\hat{V}^{(m)}_l\equiv\frac{1}{T}\int_0^T V^{(m)}\big[\hat{x}\cos (\Omega t)+\hat{p}\sin (\Omega t),t\big]e^{-il\Omega t}dt.\ \ \ 
\eea
With the harmonic of total rotating-frame Hamiltonian $\hat{H}_l=\hat{V}^{(0)}_l+\hat{V}^{(1)}_l$, the 2nd-order Floquet-Magnus expansion  ($\propto \Omega^{-2}$) is given by
\bea\label{eq-HF2}
\hat{\tilde{H}}_F^{(2)}(\hat{x},\hat{p})&=&\hat{H}_F^{(2)}(\hat{x},\hat{p})+\frac{1}{\lambda\Omega}\sum_{l\neq 0}^\infty\frac{1}{l}[\hat{V}^{(0)}_l,\hat{V}^{(1)}_{-l}]\nl
&&+\frac{1}{\lambda\Omega}\sum_{l\neq 0}^\infty\frac{1}{l}[\hat{V}^{(1)}_{-l},\hat{V}^{(0)}_{0}]e^{il\Omega t_0}.
\eea
Note that the additional harmonic $\hat{V}^{(1)}_l$ is already in the first order of perturbative parameter $\Omega^{-1}$. 
The first term on the right-hand side $\hat{H}_F^{(2)}(\hat{x},\hat{p})$ is the standard 2nd-order Flqouet-Magnus expansion term \cite{Mikami2016prb}  from the leading-order driving potential $V^{(0)}(x,t)$, see the detailed expression in Appendix~\ref{app-FM}. %
In order to mitigate the 2nd-order Floquet-Magnus expansion, we calculate the NcFT coefficient  $f^{(2)}(k,\Omega t)$ of $\tilde{H}_F^{(2)}(\hat{x},\hat{p})$, and introduce the  2nd-order driving potential
\bea
V^{(2)}(x,t)=-\int_{-\infty}^{+\infty} \frac{|k|}{2}f^{(2)}(k,\Omega t)e^{i kx}dk.
\eea
As a result, the RWA Hamiltonian of $V^{(2)}(x,t)$ will cancel the 2nd-order expansion $\tilde{H}^{(2)}_F(\hat{x},\hat{p})$. 

\begin{figure}
\centerline{\includegraphics[width=\linewidth]{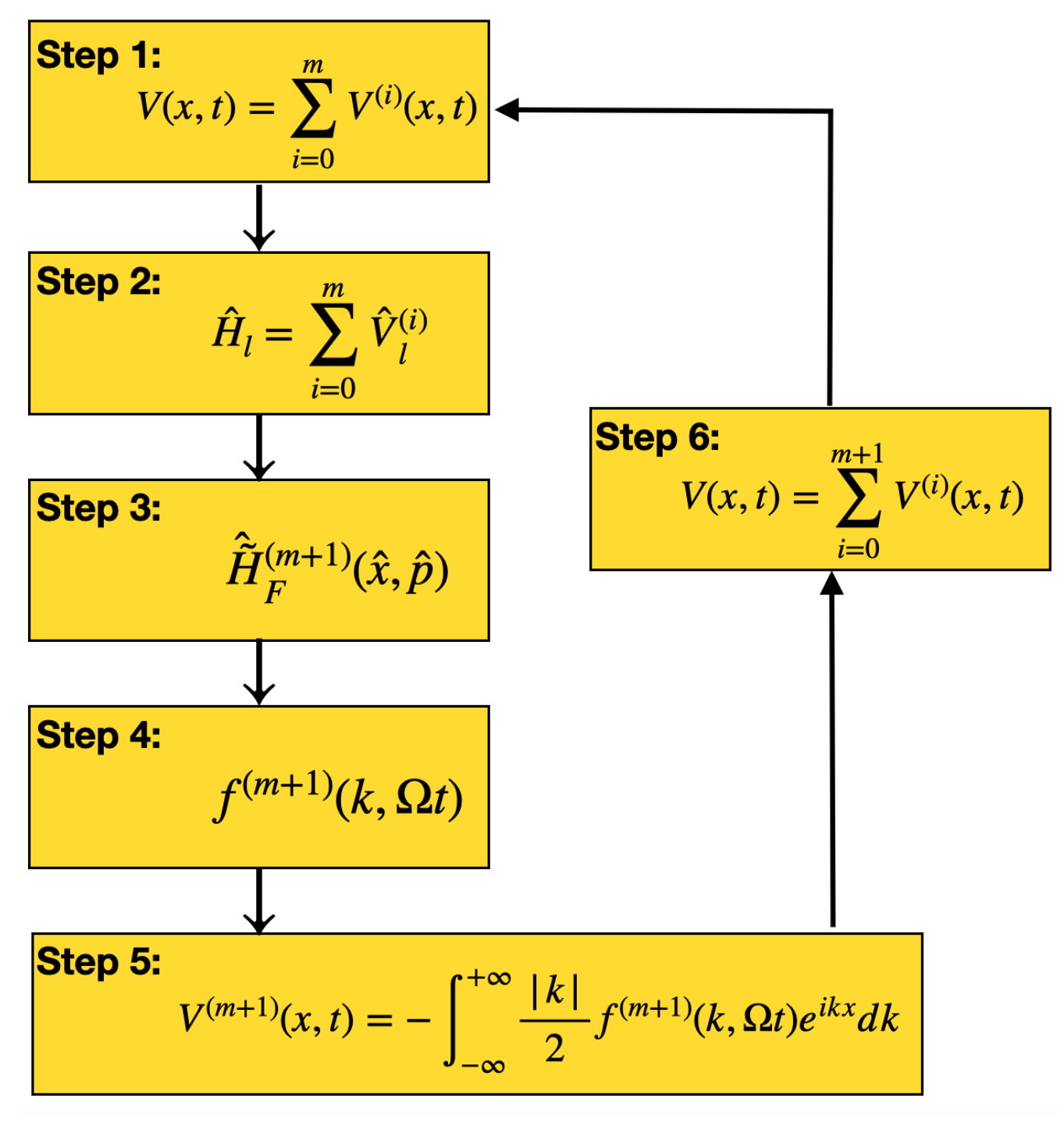}}
\caption{\label{Figures-CFE}
{\bf Perturbative framework for engineering Floquet Hamiltonian up to desired order:}  6-step procedure to mitigate high-order Floquet-Magnus terms $\hat{\tilde{H}}_F^{(m\geq1)}(\hat{x},\hat{p})$ by introducing high-order driving potentials $V^{(m\geq1)}(x,t)$, see the detailed explanation in the last paragraph of Section~\ref{sec-perfra}.
}
\end{figure}

Following the above procedure, we summarize the general perturbative framework for introducing additional driving potentials to mitigate all the high-order Floqeut-Magnus expansion terms as follows:
\begin{enumerate}
\item Summarize the driving potential up to the known order  ($\propto \Omega^{-m}$), i.e., $V(x,t)=\sum_{i=0}^mV^{(i)}(x,t)$; 
\item Construct the harmonics of Hamiltonian up to the order of $\Omega^{-m}$, i.e.,  $\hat{H}_l=\sum_{i=0}^m\hat{V}^{(i)}_l$ from Eq.~(\ref{eq-Vil});
\item Calculate the Floquet-Magnus expansion Hamiltonian up to the next order of $\Omega^{-(m+1)}$, i.e.,  $\hat{\tilde{H}}_F^{(m+1)}(\hat{x},\hat{p})$ from harmonics $\hat{H}_l=\sum_{i=0}^m\hat{V}^{(i)}_l$, by collecting all the possible terms of the order $\Omega^{-(m+1)}$ from the standard Floquet-Magnus expansion via a recursive procedure  \cite{Casas2001NJP,Blanes2009PR,Mikami2016prb};
\item  Calculate the NcFT coefficient $f^{(m+1)}(k,\Omega t)$ of $\hat{\tilde{H}}_F^{(m+1)}(\hat{x},\hat{p})$ from Eqs.~(\ref{eq-fFT}) and (\ref{eq-FTktau});
\item  Introduce the following additional driving potential
\bea\label{eq-Vix}
\ \ \ \ V^{(m+1)}(x,t)=-\int_{-\infty}^{+\infty} \frac{|k|}{2}f^{(m+1)}(k,\Omega t)e^{i kx}dk;\ \ 
\eea
\item
Update the driving potential  up to the next order ($\propto \Omega^{-(m+1)}$), i.e., $V(x,t)=\sum_{i=0}^{m+1}V^{(i)}(x,t)$.
\end{enumerate}
As sketched in Fig.~\ref{Figures-CFE}, by repeating the above six steps,  we can in principle mitigate the errors from the Floquet-Magnus expansions up to the desired order.
%

%
\subsection{Noncommutative Fourier transformation coefficients for commutators }\label{sec-ncftcom}
The higher-order Floquet-Magnus expansion terms involve commutators of harmonics $\hat{V}^{(i)}_{l}(\hat{x},\hat{p})$, that are in general complicated functions of operators $\hat{x}$ and $\hat{p}$. This makes it difficult to obtain a compact form for the Floquet Hamiltonian in practical applications. It is also impractical to calculate the NcFT coefficient of higher-order Floquet-Magnus terms directly from Eq.~(\ref{eq-fFT}) due to the difficulty of reordering operators. 

To circumvent this problem, we directly calculate the NcFT coefficient of commutators. 
Using Eqs.~(\ref{eq-Vix}) and (\ref{eq-Vil}), we define the NcFT coefficient of $\hat{V}^{(i)}_l$ by
 $$f^i_l(k,\tau)\equiv f^{(i)}(k,\tau)e^{-il\tau},$$
 where we have identified $\tau=\Omega t$. 
By assigning the NcFT coefficient $f^{0}_l(k,\tau)=f_T(k,\tau)e^{-il\tau}$ for the target Hamiltonian, the NcFT coefficient of commutator  $\hat{V}^{i,j}_{l',l''}\equiv[\hat{V}^{(i)}_{l'},\hat{V}^{(j)}_{l''}]$ is given by the transformation (see detailed derivation in Appendix~\ref{app-general})
\bea\label{eq-main-1}
f^{i,j}_{l',l''}(k,\tau)&=&\frac{ i}{2\pi}\int_0^{2\pi}\int_0^{2\pi} d\tau' d\tau''\frac{\sin\big[\lambda\frac{k''k'}{2}\sin(\tau'-\tau'')\big]}{ \big|\sin(\tau'-\tau'')\big|}
\nl
&&\times\frac{|k'k''|}{2}f^{i}_{l'}(k',\tau')f^{j}_{l''}(k'',\tau'')
\eea
and the relation
\bea\label{eq-main-2}
k'=k\frac{\sin(\tau''-\tau)}{\sin(\tau''-\tau')},\ \ \ 
k''=k\frac{\sin(\tau'-\tau)}{\sin(\tau'-\tau'')}.
\eea
%
For convenience, we define the transformation given by Eqs.~(\ref{eq-main-1})-(\ref{eq-main-2}) as a bracket operation 
\bea
f^{i,j}_{l',l''}\equiv\big\lfloor f^{i}_{l'},f^{j}_{l''}\big\rfloor
\eea
 with help of floor brackets ``$\lfloor$" and ``$\rfloor$". For more complicated commutator  $\hat{V}^{i,j,k}_{l,l',l''}\equiv[\hat{V}^{(i)}_{l},[\hat{V}^{(j)}_{l'},\hat{V}^{(k)}_{l''}]]$, the NcFT coefficient is given by 
\bea\label{eq-fijk}
f^{i,j,k}_{l,l',l''}=\big\lfloor f^{i}_{l},f^{j,k}_{l',l''}\big\rfloor=\big\lfloor f^{i}_{l},\big\lfloor f^{j}_{l'},f^{k}_{l''}\big\rfloor\big\rfloor.
\eea
The above equations reduce the calculation of commutators to the integral of c-numbers with no need for reordering operators in the target Hamiltonian.

%
\section{Application} \label{sec-app}
{The perturbative framework to mitigate higher-order Floquet-Magnus terms shown in Fig.~\ref{Figures-CFE} together with the transformation given by Eqs.~(\ref{eq-main-1})-(\ref{eq-fijk}) are the main results in this paper. In this section, we apply our perturbative method to two concrete examples. The first example is the model of a monochromatically driven oscillator, where we calculate explicitly the analytical expression for the additional driving potential up to the first-order correction. The second example is a target Hamiltonian with degenerate eigenstates of multi-component cat states, where we provide a numerically efficient procedure to calculate high-order correction drives.}

\subsection{Example I: monochromatically driven oscillator}

\subsubsection{Target Hamiltonian}
%
We consider a monochromatically driven harmonic oscillator with the following Hamiltonian
\bea\label{eq-Hcos}
\mathcal{H}(t)=\frac{1}{2}(\hat{x}^2+\hat{p}^2)+\beta\cos(\hat{x}+n\Omega t).
\eea
Such a system can be realised with a cold atom in a propagating optical lattice potential or a resonator (cavity or LC circuit ) in series with Josephson junction biased by a dc voltage \cite{Armour2013prl,Gramich2013prl,Juha2013prl,Juha2015prl,Juha2016prb,Armour2015prb,Trif2015prb,Kubala2015iop,Hofer2016prb,Dambach2017njp,Lang2021njp,Lang2022arxiv}. 
In the $n$-photon resonance condition, the system Hamiltonian (\ref{eq-Hcos}) in the rotating frame of harmonic frequency can be obtained from Eq.~(\ref{eq-Ht}), 
\bea\label{eq-Hcost}
\hat{H}(t)
&=& \beta \cos\big[\hat{p}\sin (\Omega t)+\hat{x}\cos (\Omega t)+n\Omega t\big].
\eea
From Eqs.~(\ref{eq-h0h1h3}) and (\ref{eq-Hcost}), the leading-order RWA Floquet Hamiltonian is given by \cite{Guo2016njp,Liang2018njp}
\bea\label{eq-HF0A2}
\hat{H}^{(0)}_F=\frac{\beta}{2}\Big[e^{-\frac{\lambda}{4}-i\frac{1}{2}n\pi}\Big(\frac{\lambda}{2}\Big)^{-\frac{n}{2}}\hat{a}^nL_{\hat{a}^\dagger\hat{a}}^{(-n)}({\lambda}/{2})+h.c.\Big],
\eea
where function $L_{\hat{a}^\dagger\hat{a}}^{(-n)}(\bullet)$ is the generalized Laguerre polynomials with an operator index  $\hat{a}^\dagger\hat{a}$. The above Hamiltonian Eq.~(\ref{eq-HF0A2}) is our target Hamiltonian to be engineered.
The Q-function of the target Hamiltonian is given by (see the detailed derivation in Appendix~\ref{app-QH})
\bea\label{eq-HQJ}
\langle \alpha|\hat{H}^{(0)}_F|\alpha\rangle=\beta e^{-\frac{\lambda}{4}} J_n(r)\cos(n\theta+\frac{n\pi}{2}),
\eea
where $J_n(\bullet)$ is the Bessel function of order $n$, and the parameters ($r$, $\theta$) are defined via $x=r\cos\theta$,  $p=r\sin\theta$. 

In Fig.~\ref{Fig-Hamiltonian}(a), we plot the Q-function of the target Hamiltonian (scaled by $\beta e^{-\frac{\lambda}{4}}$) in the $(x,p)$ phase space. In Fig.~\ref{Fig-Hamiltonian}(b), we show the energy spectrum of the target Hamiltonian for parameters $n=2$ and $\lambda=2.5$. In Fig.~\ref{Fig-Hamiltonian}(c), we plot the Husimi Q function of the lowest eigenenstate (quasi-ground state) of the target Hamiltonian.
%

%
\subsubsection{Symmetries and breaking}
\begin{figure}
\centerline{\includegraphics[width=\linewidth]{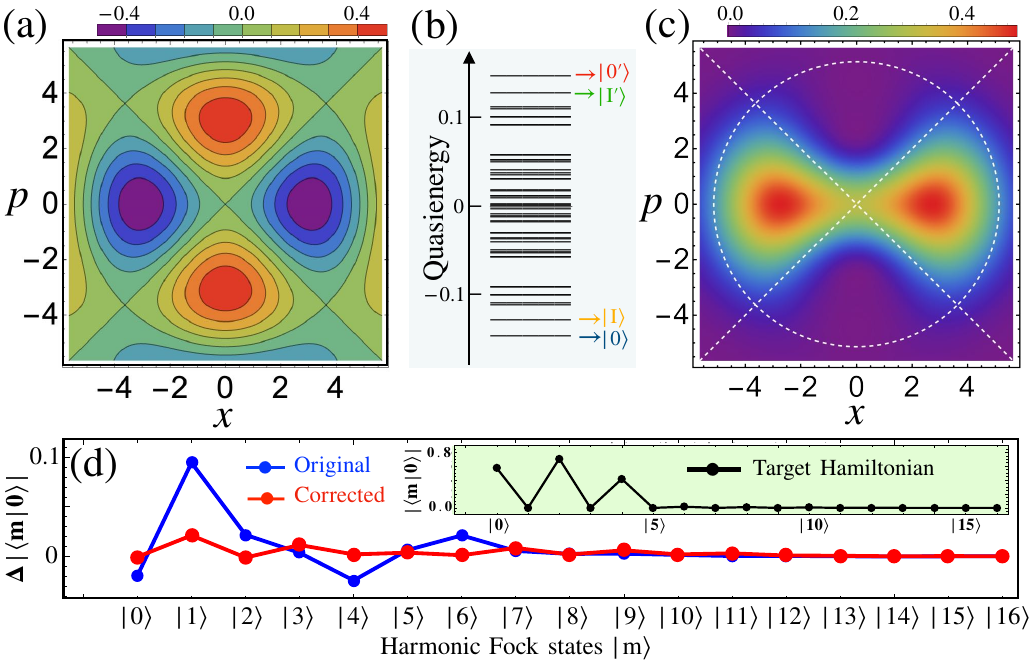}}
\caption{\label{Fig-Hamiltonian}
{\bf Target Hamiltonian.} {\bf (a)} Q function of the target Hamiltonian given by Eq.~(\ref{eq-HF0A2}) in phase space, cf. Eq.~(\ref{eq-HQJ}), scaled by factor $\beta e^{-\frac{\lambda}{4}}$ for the symmetry parameter $n=2$. {\bf (b)} Eigenspectrum of the target Hamiltonian for the parameters setting: $n=2,\ \beta=0.5$ and $\lambda=2.5$.
{\bf (c) } Husimi Q function of the quasi-ground state of the target Hamiltonian, i.e., the lowest level $|0\rangle$ marked in (b). {\bf (d) } 
Difference of the absolute probability amplitude over the harmonic Fock basis $\Delta|\langle m|0\rangle|$, cf. Eq.~(\ref{eq-dm0}), of the quasi-ground state of the original Hamiltonian Eq.~(\ref{eq-Hcos}) ({blue}), the corrected Hamiltonian up to 1st-order Floquet-Magnus expansion given by Eq.~(\ref{eq-fikl}) (red), with respect to that of the target Hamiltonian Eq.~(\ref{eq-HF0A2}) (inset, black).
}
\end{figure}

As indicated by the Q functions of target Hamiltonian $\hat{H}^{(0)}_F$ given by Eq.~(\ref{eq-HF0A2}) and quasi-ground state shown in Figs.~\ref{Fig-Hamiltonian}(a)-(c), the target Hamiltonian keeps invariant under the $n$-fold rotational operator $\hat{R}_\tau\equiv e^{-i\tau\hat{a}^\dagger\hat{a}}$ in phase space, i.e.,
\bea\label{eq-RS}
\hat{R}^\dagger_\tau\hat{H}^{(0)}_F \hat{R}_\tau=\hat{H}^{(0)}_F \ \ \mathrm{for} \ \ \tau=\frac{2\pi}{n}.
\eea
In fact, the target Hamiltonian also has the chiral symmetry that is described by \cite{Guo2016njp}
\bea
\hat{R}^\dagger_\tau\hat{H}^{(0)}_F \hat{R}_\tau=-\hat{H}^{(0)}_F \ \ \mathrm{for} \ \ \tau=\frac{\pi}{n}.
\eea
The chiral symmetry is manifested by the Q function of target Hamiltonian shown in Fig.~\ref{Fig-Hamiltonian}(a).  As a result, the quasienergy spectrum is symmetric with respect to zero as shown by Fig.~\ref{Fig-Hamiltonian}(b). 
 This $n$-fold rotational symmetry and the chiral symmetry are important for realizing bosonic codes \cite{Arne2020prx}. 
However, the above discrete rotational and chiral symmetries are obtained from the lowest-order Floquet Hamiltonian of the original Hamiltonian described by Eq.~(\ref{eq-Hcos}) in the rotating frame and thus are only valid in the RWA. The exact Floquet Hamiltonian actually does not have such symmetries.
According to Eq.~(\ref{eq-Hcost}), the discrete rotating transformation of the original Hamiltonian is given by
\bea
\hat{R}^\dagger_\tau\hat{H}(t) \hat{R}_\tau=\hat{H}(t+\tau) \ \ \mathrm{for} \ \ \tau=\frac{2\pi}{n}.
\eea
Thus the harmonics of rotated Hamiltonian are changed to be $\hat{H}_l\rightarrow \hat{H}_le^{-i\frac{2\pi l}{n}}$, cf. Eq.~(\ref{eq-Ht}). The RWA target Hamiltonian (\ref{eq-HF0A2}) that keeps the $n$-fold rotational symmetry in phase space only contains the zeroth-order harmonics $\hat{H}_{l=0}$. 
Such symmetry is deteriorated by the high-order Floquet-Magnus expansions from the harmonics $\hat{H}_{l\neq 0}$, cf., the $t_0$-dependent terms in Eq.~(\ref{eq-H1Fxp}). 
Similarly, the chiral symmetry is also broken due to  
the high-order harmonics $\hat{H}_l\rightarrow -\hat{H}_le^{-i\frac{\pi l}{n}}$. 

Our target is to protect the $n$-fold rotational symmetry and chiral symmetry by introducing additional driving potentials into the original Hamiltonian (\ref{eq-Hcos}) that can mitigate the high-order Floquet-Magnus errors.

%
\subsubsection{First-order correction drive}\label{sec-edp}
In this section, we calculate the first-order correction to the driving potential that mitigates the 1st-order Floquet-Magnus expansion $\hat{H}^{(1)}_F(\hat{x},\hat{p})$ given by Eq.~(\ref{eq-H1Fxp}).

First, by taking the driving potential in the original Hamiltonian (\ref{eq-Hcos}) as the lowest order term $V^{(0)}(x,t)=\beta\cos(\hat{x}+n\Omega t)$, the corresponding NcFT coefficient can be obtained from Eq.~(\ref{eq-Vxt-1}),
\bea
f_T(k,\Omega t)=\beta\delta(k-1)e^{in\Omega t}+A\delta(k+1)e^{-in\Omega t}.
\eea
Next, we calculate the NcFT coefficients of commutators that appear in the 1st-order Floquet-Magnus expansion Hamiltonian (\ref{eq-H1Fxp}) by identifying $\hat{H}_l=\hat{V}^{(0)}_l$. According to Eqs.~(\ref{eq-main-1}) and (\ref{eq-main-2}), we calculate analytically the NcFT coefficient of commutator $\hat{V}^{0,0}_{l,-l}=[\hat{V}^{(0)}_l,\hat{V}^{(0)}_{-l}]$ (see the detailed derivation in Appendix~\ref{app-potential})
\bea
&&f^{0,0}_{l,-l}(k,\Omega t)=\frac{ \beta^2}{\pi}\frac{\sin [\frac{\lambda}{2}\sin(2\arccos\frac{k}{2})]}{|\sin(2\arccos\frac{k}{2})|}\sin(2l\arccos\frac{k}{2})\nl
&&\times\Big[\cos(2n\Omega t)+(-1)^{n+l}\cos(2n\arccos\frac{k}{2})\Big],
\eea
and  the NcFT coefficient of  $\hat{V}^{0,0}_{-l,0}=[\hat{V}^{(0)}_{-l},\hat{V}^{(0)}_{0}]$,
\bea
&&f^{0,0}_{-l,0}(k,\Omega t)=-\frac{\beta^2}{ 2\pi }\frac{\sin [\frac{\lambda}{2}\sin(2\arccos\frac{k}{2})]}{|\sin(2\arccos\frac{k}{2})|}\nl
&&\times\Big[e^{i(2n+l){\Omega t}}\sin[l\arccos\frac{k}{2}]
-e^{-i(2n-l){\Omega t}}\sin[l\arccos(-\frac{k}{2})]\nl
&&+e^{il{\Omega t}}\Big(e^{-in\pi}\sin[(2n+l)\arccos\frac{k}{2}]\nl
&&+e^{in\pi}\sin[(2n-l)\arccos(-\frac{k}{2})]\Big)\Big].
\eea
In total, the NcFT coefficients of the 1st-order Floquet-Magnus expansion Hamiltonian (\ref{eq-H1Fxp}) is given by
\bea\label{eq-fikl}
f^{(1)}(k,\Omega t)&=&\sum_{l=1}^\infty\frac{1}{\lambda\Omega l}\Big(f^{0,0}_{l,-l}(k,\Omega t)+f^{0,0}_{-l,0}(k,\Omega t)e^{il\Omega t_0}\nl
&&+[f^{0,0}_{-l,0}(-k,\Omega t)e^{il\Omega t_0}]^*\Big).
\eea
%
 Finally, we mitigate the 1st-order Floquet-Magnus Hamiltonian~(\ref{eq-H1Fxp}) by introducing the additional engineered additional driving potential according to Eq.~(\ref{eq-V1xt}). 

\begin{figure}
\centerline{\includegraphics[width=\linewidth]{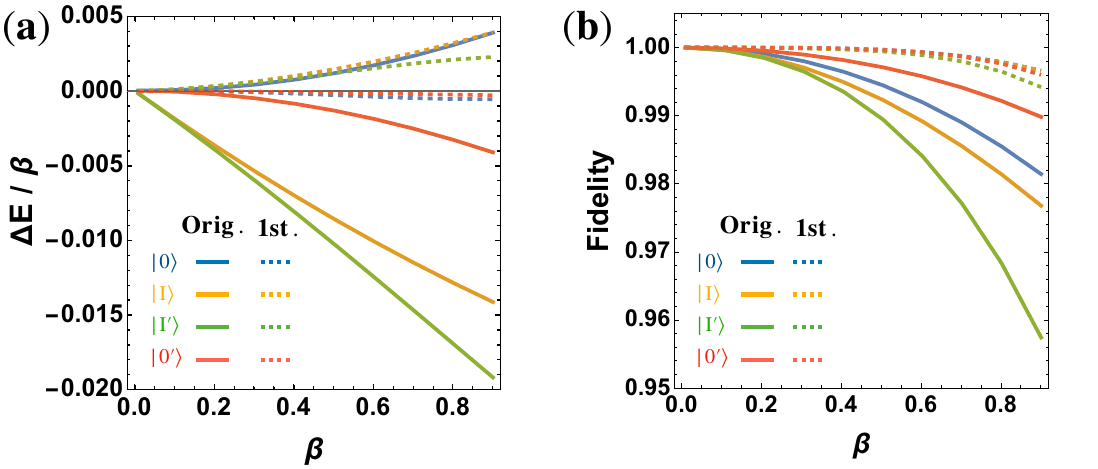}}
\caption{\label{Fig-Levels}
{\bf Engineered quasienergy spectrum and states.} 
{\bf (a)} The quasienergy level deviation $\Delta E/\beta=(E_{1st}-E_{orig})/\beta$ from the target spectrum of the original Hamiltonian (solid curves) and the 1st-order corrected Hamiltonian (dashed curves) as a function of driving amplitude $\beta$. The four colors represent the four selected quasienergy levels, cf. Fig.~\ref{Fig-Hamiltonian}(b).
 {\bf (b)} The fidelity of the four selected quasienergy states of the original Hamiltonian (solid curves) and the 1st-order corrected Hamiltonian (dashed curves)  with respect to the target states as function of driving amplitude $\beta$. Parameters setting: $n=2$, $\lambda=2.5$.
}
\end{figure}


\subsubsection{Numerical results}\label{sec-num}
We now verify our method by numerical simulations.
The eigenvalues and eigenstates of target Hamiltonian  Eq.~(\ref{eq-HF0A2})  can be directly obtained in the Fock space of harmonic oscillator $\mathbb{F}\equiv\{|m\rangle |m=0,1,\cdots \}$. To diagonalize the time-periodic Hamiltonian given by Eq.~(\ref{eq-Hcost}) and also the 1st-order corrected Hamiltonian with additional driving potential from Eqs.~(\ref{eq-V1xt}) and (\ref{eq-fikl}), we introduce the composite Hilbert space $\mathbb{F}\otimes\mathbb{T}$ that is a product of the Fock space  $\mathbb{F}$  and the temporal space $\mathbb{T}\equiv\{|e^{iM\Omega t}\rangle | M=0,\pm 1, \pm 2, \cdots\}$. In general, the eigenstate (Floquet mode) can be expressed as \cite{Grifoni1998pr}
\bea\label{eq-phit}
|\Phi_\alpha(t)\rangle=\sum_{m,M}c^{m,M}_\alpha|m\rangle\otimes |e^{iM\Omega t}\rangle.
\eea
Here, the index $\alpha$ labels the eigenlevels with quasienergy $\epsilon_\alpha$.
According to the Floquet theorem~\cite{Grifoni1998pr}, the Floquet-state solution is given by 
$
|\Psi_\alpha(t)\rangle=e^{-i\frac{\epsilon_\alpha t}{\lambda}}|\Phi_\alpha(t)\rangle.
$
In the Appnendix~\ref{app-FS}, we provide more technical details for solving the eigenproblem of the Floquet system. In the numerical simulation, a truncation of the temporal index $M$ has to be introduced. For a fixed truncation $|M_{max}|$ of temporal space, there exists an optimal truncation $l_{max}= |M_{max}|$ of harmonic index $l$ (for avoiding overcorrection). In our numerical simulations, we choose $l_{max}= |M_{max}|=10$ to obtain convergent results.

To compare the Floquet mode given by (\ref{eq-phit}) in the extended Hilbert space to the eigenstate of target Hamilton, we project the Floquet mode onto the harmonic Fock basis, i.e., $|\Phi_\alpha(t)\rangle=\sum_{m}\Phi^m_\alpha(t)|m\rangle$ with the probability amplitude  on the Fock state $|m\rangle\ (m=0,1,\cdots)$ given by
\bea\label{eq-phim}
\Phi^m_\alpha(t)\equiv\big(\langle t|\otimes\langle m|\big)|\Phi_\alpha(t)\rangle=\sum_{M}c^{m,M}_\alpha e^{iM\Omega t}.
\eea
It is clear that the probability amplitude of Floquet mode on the Fock basis is time-\textit{dependent} with period $2\pi/\Omega$. The periodic time dependence of the Floquet modes describes the so-called \textit{micromotion}. The stroboscopic dynamics of Hamiltonian (\ref{eq-Hcost}) depends on the choice of initial time $t_0$. 

For simplicity, we first consider the initial reference time choice of $t_0=0$.
In Fig.~\ref{Fig-Hamiltonian}(d), we compare the absolute probability amplitude $|\langle m|0\rangle|$ over the harmonic Fock states for the quasi-ground state of the target Hamiltonian, the original Hamiltonian and the 1st-order corrected Hamiltonian, by plotting their difference 
\bea\label{eq-dm0}
\Delta|\langle m|0\rangle|\equiv |\langle m|0_{orig/1st}\rangle|-|\langle m|0\rangle|.
\eea
It is clearly shown that the correction with an additional driving field (red dots) makes the state much closer to the target state than that without correction ({blue dots}).

In Fig.~\ref{Fig-Levels}(a), we compare the errors of the selected quasienergy levels of target Hamiltonian, i.e., the upper two and lower two levels marked in Fig.~\ref{Fig-Hamiltonian}(b),  from the original Hamiltonian (black) and the 1st-order corrected Hamiltonian (red) as a function of driving amplitude. Both errors from the original Hamiltonian and the 1st-order corrected Hamiltonian decrease as the driving amplitude approaches zero. For every selected level, the 1st-order correction indeed reduces the errors. The figure also verifies the fact that the high-order Floquet-Magnus expansion terms destroy the chiral symmetry as the quasienergy corrections to level pairs $|0\rangle$, $|0'\rangle$ (and $|\mathrm{I}\rangle$, $\mathrm{|I'}\rangle$) are not symmetric with respect to zero.

\begin{figure}
\centerline{\includegraphics[width=\linewidth]{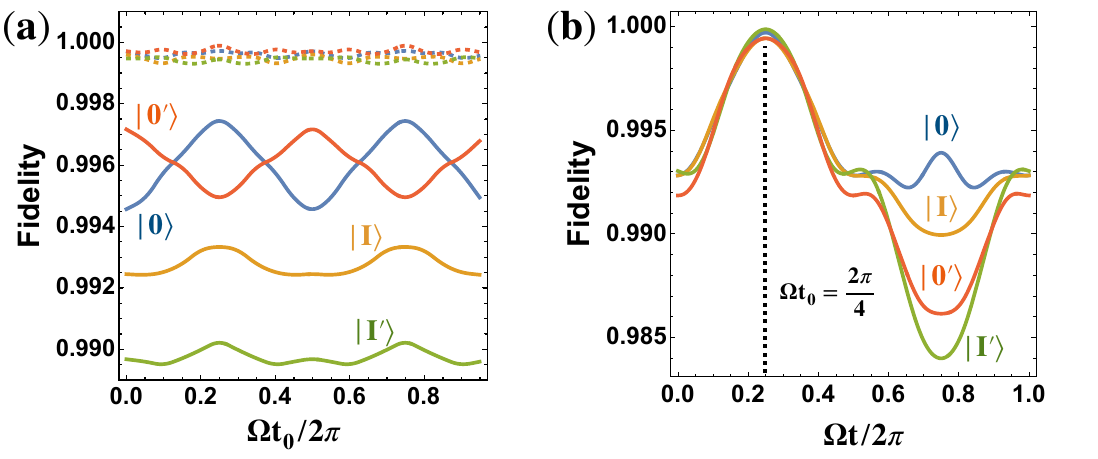}}
\caption{\label{Fig-Drivephase}
{\bf Dependence of initial reference time.} 
{\bf (a)} The fidelity of the four selected quasienergy states of the original Hamiltonian (solid curves) and the 1st-order corrected Hamiltonian (dashed curves)  with respect to the target states as a function of initial reference time $t_0$, which is set to define the Floquet Hamiltonian in Eq.~(\ref{eq-h0h1h3}).
{\bf (b)} For fixed initial reference time $t_0=2\pi/(4\Omega)$,  the fidelity of four engineered quasienergy eigenstates from the 1st-order corrected Hamiltonian with respect to the target states as a function of evolution time (micromotion) in one Floquet period.  Parameters setting: $n=2$, $\lambda=2.5$ and $\beta=0.5$.
}
\end{figure}

 In Fig.~\ref{Fig-Levels}(b), we compare the fidelity (defined as the absolute value of the inner product for two pure states \cite{james2001pra}) of the selected quasienergy levels of the original Hamiltonian and the 1st-order corrected Hamiltonian with respect to the target Hamiltonian as a function of driving amplitude. Both the fidelities of states from the original lab-frame Hamiltonian and the 1st-order corrected Hamiltonian increase as the driving amplitude approaches zero, and the 1st-order corrected Hamiltonian results in higher fidelity than that without correction.

Now we continue to investigate the dependence of engineered Hamiltonian on the choice of initial reference time $t_0$ in Eq.~(\ref{eq-HFt0}), or equivalently, setting the initial driving phase in Eq.~(\ref{eq-Hcos}). Obviously from Eq.~(\ref{eq-h0h1h3}), the RWA Floquet Hamiltonian does not depend on the choice of $t_0$. 
But according to the exact definition in Eq.~(\ref{eq-HFt0}), the Floquet Hamiltonian is a function of initial time $t_0$. In Fig.~\ref{Fig-Drivephase}(a), we plot the fidelity of four selected quasienergy levels of the original Hamiltonian (solid curves) with respect to that of the target Hamiltonian as a function of initial reference time $t_0$. The change of fidelity as a function of initial time $t_0$ comes from the high-order Magnus expansions of the Floquet Hamiltonian. By introducing the 1st-order correction driving field according to Eq.~(\ref{eq-fikl}) as a function of $t_0$, the fidelity deviation is much suppressed as shown by the dashed curves in Fig.~\ref{Fig-Drivephase}(a).

As discussed above, the time evolution of probability amplitude on the Fock basis given by Eq.~(\ref{eq-phim})  reflects the micromotion of Floquet mode. The Floquet Hamiltonian $\hat{H}_F(t_0)$ describes the stroboscopic dynamics in the lab frame starting from initial time $t_0$, i.e., the micromotion of Floquet mode  Eq.~(\ref{eq-phit})  at stroboscopic time steps $t=nT+t_0$ ($n\in \mathbb{Z}^+$). 
For a fixed initial time $t_0=2\pi/(4\Omega)$, we plot in Fig.~\ref{Fig-Drivephase}(b) the fidelity of four selected quasienergy levels of target Hamiltonian with respect to the time-evolution Floquet modes of the 1st-order corrected Hamiltonian according to Eq.~(\ref{eq-phim}). Clearly, the fidelity of all the four selected quasienergy levels reaches a maximum when the evolution time $t$ in Eq.~(\ref{eq-phim}) coincides with the choice of initial reference time $t_0$ in Eq.~(\ref{eq-HFt0}). The maximum of fidelity points out which micromotion state of the 1st-order corrected Hamiltonian represents the stroboscopic dynamics described by the target Hamiltonian.

{ 
\subsection{Example II: multi-component cat states} 

In the first example, the monochromatic drive allows us to calculate explicitly the analytical expression for the additional driving potential up to the first-order correction.  However, it is generally impossible to obtain a compact analytical result using the transformation formula given by Eqs.~(\ref{eq-main-1})-(\ref{eq-fijk}). It is also numerically time-consuming to directly calculate the double integrals in Eq.~(\ref{eq-main-1}). In the second example, we provide a numerically efficient procedure to calculate high-order correction drives for a general target Hamiltonian and engineer a target Hamiltonian with degenerate eigenstates of multi-component cat states that are important for fault-tolerant bosonic quantum computing.

\subsubsection{Target Hamiltonian}

The target Hamiltonian we aim to engineer is a general $q$-fold discrete rotational lattice in phase space given by
\bea\label{eq-sm-HTq}
\hat{H}_{T}=\frac{\beta}{|\alpha_0|^{2q}}e^{-\gamma\hat{a}^\dagger\hat{a}}(\hat{a}^{\dagger q}-\alpha_0^{*q} )(\hat{a}^{q}-\alpha_0^{q} )e^{-\gamma\hat{a}^\dagger\hat{a}}.
\eea
Here, the factor $e^{-\gamma\hat{a}^\dagger\hat{a}}$ with $\gamma>0$ is introduced to suppress the divergence in calculating NcFT coefficient. 
Using the identity $e^{-\gamma\hat{a}^\dagger\hat{a}}|\alpha\rangle=e^{-\frac{1}{2}(1-e^{-2\gamma})|\alpha|^2}|\alpha e^{-\gamma}\rangle,$ we have the Hamiltonian Q-function as follows 
\bea\label{eq-sm-HTxp-q}
H^{(Q)}_T&\equiv&\langle \alpha|\hat{H}_T|\alpha\rangle\\
&=&\frac{\beta}{|\alpha_0e^{\gamma}|^{2q}}\exp(-\frac{x^2+p^2}{2\lambda\sigma^2_\gamma})\Big|\Big(\frac{x+ip}{\sqrt{2\lambda}}\Big)^q-\alpha^q_0e^{q\gamma}\Big|^2\nn
\eea
with the parameter $\sigma_\gamma=1/\sqrt{1-e^{-2\gamma}}$. In Fig.~\ref{Fig-Cat}(a), we plot the Q-function of the four-fold ($q=4$) symmetric target Hamiltonian $H^{(Q)}_{T}(x,p)/\beta$, where the phase-space coordinates have been rescaled such that the $q$ global minima fulfill $|x+ip|=1$. 
Remarkably, the $q$ standard coherent states $|\alpha_0 e^{\gamma+i\frac{2\pi s}{q}}\rangle$ with $s=0,1,\cdots,q-1$ are the degenerate exact zero-energy eigenstates of Hamiltonian~(\ref{eq-sm-HTq}), which means quantum fluctuations do not introduce any tunneling between these coherent states. 

\subsubsection{Bosonic code states}

According to the Bloch theorem extended in phase space \cite{Guo2013prl,Arne2020prx}, the $q$-fold rotational symmetric Hamiltonian has eigenstates in the Bloch form of 
\bea\label{eq-sm-psilm}
|\psi_{l,s}\rangle=\frac{1}{\sqrt{\mathcal{N}_{l,s}}}\sum_{p=0}^{q-1}e^{isp\frac{2\pi}{q} }\Big(\hat{R}^{\dagger}_{\frac{2\pi}{q}}\Big)^p|\phi_l\rangle
\eea
with $\hat{R}_{2\pi/q}=e^{-i\frac{2\pi}{q}\hat{a}^\dagger\hat{a}}$ the rotational operator, cf. Eq.~(\ref{eq-RS}). Here, $l$ is the index of the Bloch bands,  $s$ is the quasinumber representing the generalized parity of state, $|\phi_l\rangle$ is the cell state of $l$-th Bloch band and $\mathcal{N}_{l,s}$ is the normalized factor. For the target Hamiltonian~(\ref{eq-sm-HTq}),  the cell state for the lowest band  is  the coherent state $|\alpha_0 e^{\gamma}\rangle$, and the Bloch states are  
\bea\label{eq-sm-psimcode}
|\psi_s\rangle=\frac{1}{\sqrt{\mathcal{N}_s}}\sum_{p=0}^{q-1}e^{is\frac{2\pi p}{q} }|\alpha_0 e^{\gamma+i\frac{2\pi p}{q}}\rangle.
\eea
Here, we have omitted the band index $l=0$ for simplicity. The above $q$ degenerate $q$-component cat states construct the $q$-dimensional ground state manifold of the target Hamiltonian. They are classified as the rotational bosonic codes that are important for fault-tolerant bosonic quantum computing \cite{Arne2020prx}. Importantly, the cat states given by Eq.~(\ref{eq-sm-psimcode}) offer the further advantage of automatic quantum error correction against photon losses by tracking the results of parity measurements and updating the knowledge on the code basis without feedback operations \cite{Gertler2021nature,guo2024arxiv}.

\subsubsection{Engineering driving potential}

The analytical expression for the NcFT coefficient of the target Hamiltonian (\ref{eq-sm-HTq}) (scaled by $\beta$) is given by \cite{Guo2013prl}
\bea\label{eq-fT-n}
&&f_T(k,\tau)=\frac{\lambda e^{\frac{\lambda}{4}k^2} \sigma^{2(q+1)}_\gamma}{|\alpha_0e^{\gamma}|^{2q}}\Big[n!{_1F_1}(1+q;1;-\frac{\lambda}{2}\sigma^2_\gamma k^2)\nl
&&-\Big(-ie^{-i\tau}\alpha_0e^{\gamma}\sqrt{\frac{\lambda}{2}}\Big)^qk^q{_1F_1}(1+q;1+q;-\frac{\lambda}{2}\sigma^2_\gamma k^2)\nl
&&-\Big(-ie^{i\tau}\alpha_0^*e^{\gamma}\sqrt{\frac{\lambda}{2}}\Big)^qk^q{_1F_1}(1+q;1+q;-\frac{\lambda}{2}\sigma^2_\gamma k^2)\nl
&&+\Big|\frac{\alpha_0e^{\gamma}}{\sigma_\gamma}\Big|^{2q}{_1F_1}(1;1;-\frac{\lambda}{2}\sigma^2_\gamma k^2)\Big],
\eea
where $\tau=\Omega t\in [0,2\pi)$ is the dimensionless time parameter, $J_q(\bullet)$ is the Bessel function and  ${_1F_1}(a;b;\bullet)$ is the Kummer confluent hypergeometric function. The zeroth-order real-space driving potential $V^{(0)}(x,t)$ can be directly engineered from Eq.~(\ref{eq-Vxt-1}).
In order to calculate the first-order NcFT coefficient $f^{(1)}(k,\tau)$ that is needed to construct the first-order correction drive $V^{(1)}(x,t)$, cf. Eq.~(\ref{eq-V1xt}), one can again apply the transformation formula given by Eqs.~(\ref{eq-main-1})-(\ref{eq-fijk}). However, different from the monochromatically driven harmonic oscillator, it is impossible to obtain explicitly the analytical expression for the correction drives in this case. It is also numerically time-consuming to directly calculate the double integrals in Eq.~(\ref{eq-main-1}). 

Here, we provide a numerically efficient procedure to calculate high-order correction drives.
The first step is to calculate the matrix of the zeroth-order drive of the rotating frame in the Fock representation, cf. Eq.~(\ref{eq-Vxt-1}),
\bea\label{eq-s1}
V^{(0)}_{nm}(\tau)&=&\int_{-\infty}^{+\infty}\frac{|k|}{2}f_T(k, \tau)\langle n|e^{i k[\hat{x}\cos \tau+\hat{p}\sin\tau)}|m\rangle dk,\nl
\eea
with the matrix element given by
\bea\label{eq-s2}
&&\langle n|e^{ik[\hat{x}\cos\tau+\hat{p}\sin\tau]}|m\rangle\\
&&=e^{-\frac{\lambda}{4}k^2-i(m-n)\tau}\Big(ik\sqrt{\frac{\lambda}{2}} \ \Big)^{m-n}\sqrt{\frac{n!}{m!}}L^{m-n}_n\Big(\frac{\lambda}{2}k^2\Big).\nn
\eea
The second step is to calculate the harmonics of the zeroth-order drive of the rotating frame in the Fock representation, cf. Eq.~(\ref{eq-Vxt-1}), by
\bea\label{eq-s3}
\hat{V}^{(0)}_{l,nm}=\frac{1}{2\pi}\int_0^{2\pi}V^{(0)}_{nm}(\tau)e^{-il\tau} d\tau.
\eea
The third step is to calculate the matrix of the first-order Magnus expansion term $\hat{H}^{(1)}_F$ using the above harmonics, cf. Eq.~(\ref{eq-Vxt-1}), which in the Fock representation is 
\bea\label{eq-s4}
\hat{H}^{(1)}_F=\sum_{n,m}c_{nm}|n\rangle\langle m|.
\eea
The fourth step is to calculate the NcFT coefficient of the first-order Magnus expansion Hamiltonian $ H^{(1)}_F(\hat{x},\hat{p})$ 
\bea\label{eq-s5}
f^{(1)}(k,\tau)=\sum_{n,m}c_{nm} f_{nm}(k,\tau),
\eea
where $f_{nm}(k,\tau)$ is  the NcFT coefficient of the operator $|n\rangle\langle m|$ given by \cite{Guo2013prl}
\bea\label{eq-sm-fnm}
f_{nm}(k,\tau)
&=&e^{\frac{\lambda}{4}k^2}\sqrt{\frac{n!}{m!}}\Big(ie^{i\tau}\frac{1}{k}\sqrt{\frac{2}{\lambda}}\Big)^{m-n}\frac{\lambda}{\Gamma(1-m+n)}\nl
&&\times {_1F_1}(1+n;1-m+n;-\frac{\lambda}{2}k^2).
\eea
Here, ${_1F_1}(a;b;z)$ is the Kummer confluent hypergeometric function.  Finally, the 1st-order Floquet-Magnus Hamiltonian~(\ref{eq-H1Fxp}) can be mitigated by engineering additional driving potential $V^{(1)}(x,t)$ from Eq.~(\ref{eq-V1xt}). The higher-order correction dives can be calculated by repeating Eqs.~(\ref{eq-s1})-(\ref{eq-sm-fnm}) together with higher-order Magnus expansions, cf., the 2nd-order Magnus term Eq.~(\ref{eq-HF2}).

In fact, we can start from Eq.~(\ref{eq-s4}) by expressing an abitrary target Hamiltonian as $\hat{H}_T=\sum_{n,m}c^T_{nm}|n\rangle\langle m|$, where coefficients $c^T_{nm}$ can be random numbers. Then, we calculate the NcFT coefficient $f_T(k,\tau)$ from Eqs.~(\ref{eq-s5}) and (\ref{eq-sm-fnm}). By repeating the steps according to Eqs.~(\ref{eq-s1})-(\ref{eq-sm-fnm}), we can calculate the NcFT coefficients for higher-order correction dives and mitigate non-RWA errors.

\begin{figure}
\centerline{\includegraphics[width=\linewidth]{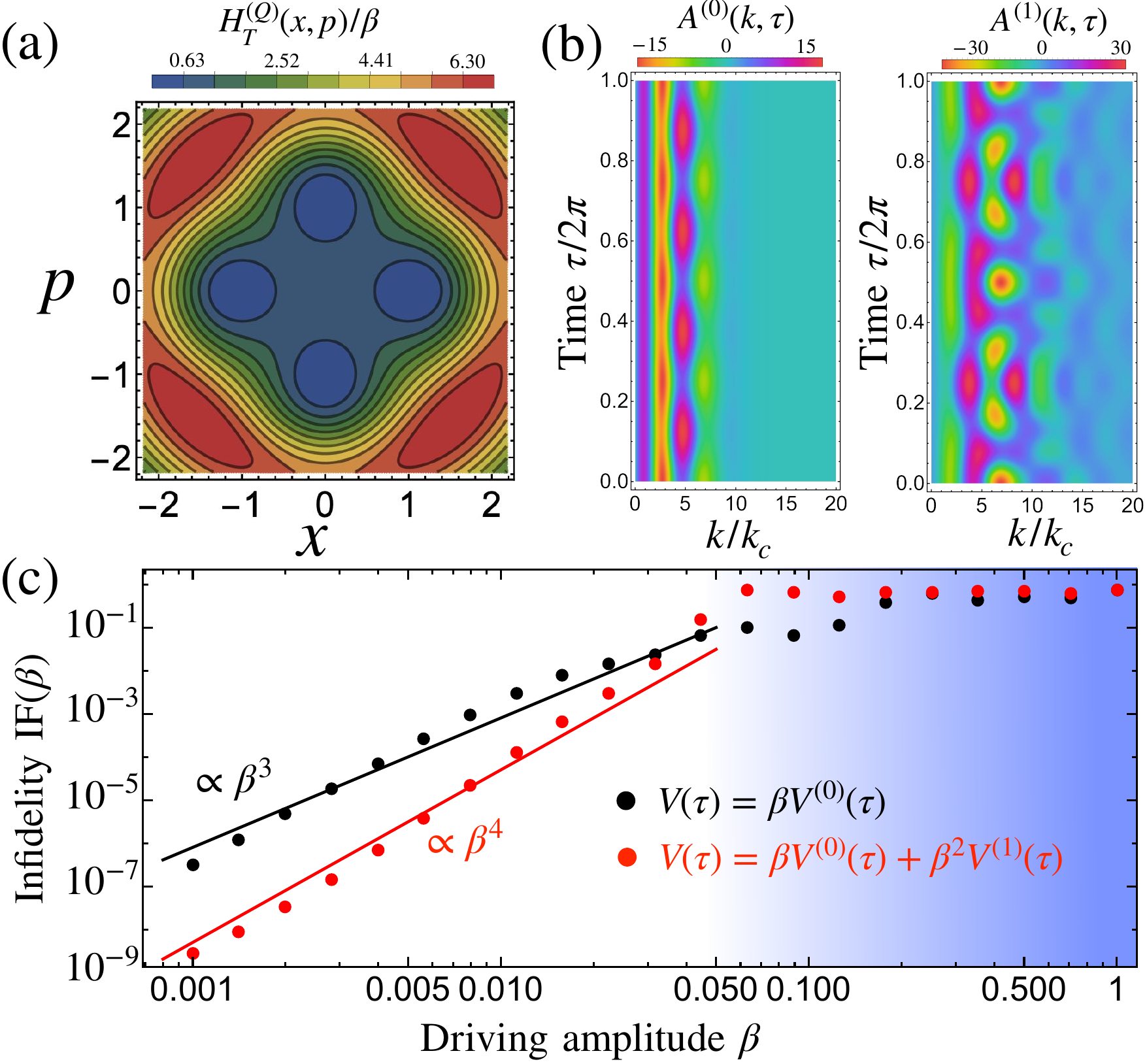}}
\caption{\label{Fig-Cat}{
{\bf Engineering four-fold symmetric Hamiltonian.} {\bf (a)} Q-function of the target Hamiltonian~(\ref{eq-sm-HTq}) in phase space $H^{(Q)}_T(x,p)/\beta$, cf. Eq.~(\ref{eq-sm-HTxp-q}), with parameters: $q=4$, $\lambda=1/4$, $\alpha_0=1.198$ and $\gamma=1/4$. {\bf (b)}  Charts of the zeroth-order drive $A^{(0)}(k,\tau)\equiv kf_T(k,\tau)/\beta$ (left), and the first-order correction drive $A^{(1)}(k,\tau)\equiv kf^{(1)}(k,\tau)/\beta^2$ (right), as functions of wavenumber $k$ divided by $k_c=\sqrt{(1-e^{-2\gamma})/4\lambda}$ and dimensionless time $\tau=\Omega t$.
{\bf (c) } Log-log plot of the infidelity $\mathrm{IF}(\beta)\equiv 1-F(\beta)$ as a function of driving amplitude $\beta$ for the zeroth-order drive $V(\tau)=\beta V^{(0)}(\tau)$ (black dots) and the drive with first-order correction $V(\tau)=\beta V^{(0)}(\tau)+\beta^2V^{(1)}(\tau)$ (red dots). The black and red lines indicate the power-law scaling of infidelity $\mathrm{IF}(\beta)\propto\beta^3$ and $\mathrm{IF}(\beta)\propto\beta^4$ respectively. }
}
\end{figure}

\subsubsection{Numerical results}

We now show the numerical results for engineering target Hamiltonian Eq.~(\ref{eq-sm-HTq}) with four-fold rotational symmetry ($q=4$). In this case, the target Hamiltonian Eq.~(\ref{eq-sm-HTq}) has four 4-component cat states given by Eq.~(\ref{eq-sm-psimcode}). By assigning coherent number $\alpha=\alpha_0e^\gamma\in\mathbb{R}$, two of the four cat states are used as code states 
\bea\label{eq-sm-4cat}
 \left \{ \begin{array}{lll}
|\psi_{s=0}\rangle\equiv\frac{1}{\sqrt{\mathcal{N}_0}}\big(|\alpha\rangle+|-\alpha\rangle+|i\alpha\rangle+|-i\alpha\rangle\big)\label{eq-sm-4cat1}\\
|\psi_{s=2}\rangle\equiv\frac{1}{\sqrt{\mathcal{N}_2}}\big(|\alpha\rangle+|-\alpha\rangle-|i\alpha\rangle-|-i\alpha\rangle\big)\label{eq-sm-4cat2},  
\end{array} \right.
\eea
to encode quantum information~\cite{Mirrahimi2014njp,Leghtas2013prl,Terhal2020iop}, and the other two are used as error states
\bea\label{eq-sm-4cat-erorr}
 \left \{ \begin{array}{lll}
|\psi_{s=1}\rangle\equiv\frac{1}{\sqrt{\mathcal{N}_{1}}}\big(|\alpha\rangle-|-\alpha\rangle-i|i\alpha\rangle+i|-i\alpha\rangle\big),\label{eq-sm-4cat1-e}\\
|\psi_{s=3}\rangle\equiv\frac{1}{\sqrt{\mathcal{N}_{3}}}\big(|\alpha\rangle-|-\alpha\rangle+i|i\alpha\rangle-i|-i\alpha\rangle\big)\label{eq-sm-4cat2-e}  
\end{array} \right.
\eea
to detect errors. Here, the normalization factors are given by $\mathcal{N}_{s}=8e^{-\alpha^2}(\cosh\alpha^2+(-1)^{\frac{s}{2}}\cos\alpha^2)$ for $s=0, 2$ and $\mathcal{N}_{s}=8e^{-\alpha^2}[\sinh\alpha^2+(-1)^{\frac{s-1}{2}}\sin\alpha^2]$ for $s=1, 3$. The code and error cat states are the eigenstates of the photon parity operator $\hat{\prod}\equiv e^{i\pi\hat{a}^\dagger\hat{a}}$ with eigenvalues ``+1" and ``-1" respectively.  The single photon loss changes the parity and thus can be detected by parity measurement \cite{haroche2007,sun2014nat,rosenblum2018sci}. However, to ideally correct the single-photon loss error, the code cat states given by Eq.~(\ref{eq-sm-4cat}) need to satisfy the Knill-Laflamme condition \cite{chuang2010book,Terhal2020iop}: $\tan\alpha^2=-\tanh\alpha^2.$ The discrete values of coherent number $\alpha$ that satisfy the quantum error condition are called ``sweet spots". Here, we choose the smallest sweet spot $\alpha\approx 1.538$ as an instance.

 In Fig.~\ref{Fig-Cat}(b), we plot the charts of the zeroth-order drive  $A^{(0)}(k,\tau)\equiv kf_T(k,\tau)/\beta$, cf. Eq.~(\ref{eq-Vxt-1}), and the first-order correction drive $A^{(1)}(k,\tau)\equiv kf^{(1)}(k,\tau)/\beta^2$, cf. Eq.~(\ref{eq-V1xt}), as functions of wavenumber $k$ and time $\tau$. In our case, both $A^{(0)}(k,\tau)$  and $A^{(1)}(k,\tau)$ are real numbers.
 To quantify the approximation of the Flouqet Hamiltonian generated by the engineered drives to the target Hamiltonian, we define the fidelity of the Floquet time-evolution operator, cf. Eq.~(\ref{eq-HFt0}), on the four cat states by
\bea\label{eq-Fidelity}
F(\beta)=\frac{1}{4}\sum^3_{m=0}\Big |\Big\langle\psi_m\Big |\mathcal{T}\exp[-i\frac{1}{\lambda}\int_{0}^{2\pi}\hat{H}(\tau)d\tau\Big|\psi_{m}\Big\rangle\Big |,\  \ 
\eea
where $\hat{H}(\tau)= V(\hat{x}\cos \tau+\hat{p}\sin \tau)$ is the engineered Hamiltonian in the rotating frame, cf. Eq.~(\ref{eq-Ht}).  

In Fig.~\ref{Fig-Cat}(c), we show the log-log plot of the infidelity $\mathrm{IF}(\beta)\equiv 1-F(\beta)$ as a function of driving amplitude $\beta$ for the zeroth-order drive $V(\tau)=\beta V^{(0)}(\tau)$ and the drive with first-order correction $V(\tau)=\beta V^{(0)}(\tau)+\beta^2V^{(1)}(\tau)$. Clearly, the infidelity exhibits a power-law scaling behavior in the weak driving limit. For the zeroth-order drive, the Flquet Hamiltonian only approximates the target Hamiltonian up to the order of $\beta$, i.e.,
\bea\label{eq-erorr-0}
\mathcal{T}e^{-i\frac{1}{\lambda}\int_{0}^{2\pi}\hat{H}(\tau)d\tau}&=&e^{-i\frac{2\pi}{\lambda}\beta\hat{H}^{(0)}_F-i\frac{2\pi}{\lambda}\beta^2\big(\sum_n\beta^{n-1}\hat{H}^{(n\geq 1)}_F\big)}\nl
&=&e^{-i\frac{2\pi}{\lambda}\hat{H}_T}e^{\beta^3\hat{O}_{erorr}}\nl
&\approx&e^{-i\frac{2\pi}{\lambda}\hat{H}_T}(1+\beta^3\hat{O}_{erorr})
\eea
where we have used $\beta\hat{H}^{(0)}_F=\hat{H}_T$ and introduced the error operacotr $\hat{O}_{erorr}$ from higher-order Floquet-Magnus expansion. In the last equality, we have adapted the Taylor expansion up to the first order for $\beta\rightarrow 0$. By taking Eq.~(\ref{eq-erorr-0}) into the fidelity Eq.~(\ref{eq-Fidelity}), we obtain the scaling behavior of the infidelity $\mathrm{IF}(\beta)\propto\beta^3$.
For the drive with 1st-order correction, we have the scaling behavior of the infidelity $\mathrm{IF}(\beta)\propto\beta^4$ due to the mitigation of $\beta^2\hat{H}^{(1)}_F$ by the first-order correction drive $\beta V^{(1)}(\tau)$. This scaling behavior analysis is verified by our numerical results shown in  Fig.~\ref{Fig-Cat}(c).

}

%
\section{Discussions and outlooks}\label{sec-discussion}

%
\subsection{Effective Hamiltonian}\label{sec-heff}
We further elucidate the subtle $t_0$-dependence in the high-order Magnus expansion of Floquet Hamiltonian, cf., Eq.~(\ref{eq-H1Fxp}). This seems inconsistent with the independence of quasienergies on the choice of $t_0$ (i.e., the initial phase of driving potential) from Floquet theory \cite{Floquet1883,Shirley1965pr,Sambe1973pra,Grifoni1998pr,Eckardt2015NJP}.
 It was argued that the $t_0$-dependence of quasienergy spectrum is spurious in the sense that the $t_0$-dependent terms in the $m$-th order ($\propto \Omega^{-m}$) Floquet-Magnus expansion will not cause changes of the spectrum within the $m$-th order but contribute to the next-order ($\propto\Omega^{-m-1}$) correction of the quasienergy spectrum \cite{Eckardt2015NJP}.   

 It has been known that the $t_0$-dependence of the Floquet-Magnus expansion can be removed by a proper gauge transformation \cite{Eckardt2015NJP,Mikami2016prb}. In fact,  the time evolution operator in one period can be written with an effective Hamiltonian operator  \cite{Mikami2016prb,marin2015aip}
\bea
U(t_0+T,t_0)=e^{-i\hat{\Lambda}(t_0)}e^{-i\frac{T}{\lambda}\hat{F}}e^{i\hat{\Lambda}(t_0)}.
\eea
Here, the time-\textit{independent} operator $\hat{F}$ is defined as the \textit{effective Hamiltonian}, and the temporal periodic operator $\hat{\Lambda}(t)=\hat{\Lambda}(t+T)$ is the so-called \textit{micromotion operator.} 
From Eq.~(\ref{eq-HFt0}), the effective Hamiltonian is related to Floquet Hamiltonian via 
\bea\label{eq-FT}
\hat{F}=e^{i\hat{\Lambda}(t_0)}\hat{H}_Fe^{-i\hat{\Lambda}(t_0)}.
\eea
If the gauge condition $\hat{\Lambda}(t_0)=0$ is chosen, we arrive at the Floquet-Magnus expansions discussed in this paper, and the corresponding effective Hamiltonian $\hat{F}$ becomes the Floquet Hamiltonian $\hat{H}_F$. If the gauge condition $\int_0^T\hat{\Lambda}(t)dt=0$ is chosen, we remove the $t_0$-dependent terms in the Floquet-Magnus expansions and arrive at the van Vleck degenerate perturbation theory \cite{Casas2001NJP}.

According to Eqs.~(\ref{eq-HFt0}) and (\ref{eq-FT}), the effective Hamiltonian $\hat{F}(\hat{x},\hat{p})$ does not describe the stroboscopic dynamics of Hamiltonian $H(t)$ but the transformed Hamiltonian $\hat{H}_{\Lambda}(t)=e^{i\hat{\Lambda}(t_0)}\hat{H}(t)e^{-i\hat{\Lambda}(t_0)}$. 
As a consequence, even the engineered effective Hamiltonian $\hat{F}$ is $n$-fold rotational symmetric in phase space, the direct stroboscopic state from the system Hamiltonian (described by Floquet Hamiltonian $\hat{H}_F$) does not has such symmetry in general.
In principle, if we engineer the Hamiltonian directly with the form of $\hat{H}_{\Lambda}(t)$, then the stroboscopic dynamics is described by the $n$-fold rotational symmetric Hamiltonian $F(\hat{x},\hat{p})$. However, the engineered driving potential $V(t)$ could be a complicated function of momentum operator $\hat{p}$ as the micromotion operator $\hat{\Lambda}(t_0)$ is also a complex function of position and momentum operators \cite{Mikami2016prb}. This is not realistic because the driving potential is only a function of the position in the laboratory frame.

\subsection{Bosonic code state preparation} \label{sec-bcsp}
As mentioned in Section~\ref{sec-intro},  our APSHE method combined with the adiabatic ramp protocol~\cite{xanda2023arxiv} can be exploited to prepare a desired quantum bosonic code state, i.e., Schr\"odinger-cat state or binomial code state. In fact, in our previous work~\cite{guo2024prl}, we have demonstrated the preparation of a multicomponent cat state in the ground state manifold of a properly designed Hamiltonian based on RWA. Although the preparation is against the noisy effects of dissipation and dephasing,  it remained a problem how to mitigate the errors from high-order Floquet-Magnus expansion terms. The present work provides a perturbative solution for this problem. The desired symmetries of the target Hamiltonian are protected by the systematic construction of additional driving potentials. The non-RWA deviation could be reduced by fine-tuning the driving potential to account for higher-order Floquet-Magnus expansion terms.
As a result, our method provides a general protocol to generate arbitrary \textit{nonlinear} transformation between bosonic states. Previously, the arbitrary \textit{linear} bosonic transformation has been proposed by Xiang et al. \cite{xiang2023prl}.

Furthermore, we emphasize that our method can synthesize arbitrary Hamiltonian even without any phase-space symmetries, i.e., a potential well with a sharp boundary that would lead to topologically robust edge transport due to the noncommutative nature of phase space~\cite{guo2024prl}. 

\subsection{Experimental implementations}  \label{sec-ei}
According to Eq.~(\ref{eq-Vxt-2}), to design arbitrary Hamiltonians in phase space, we need the ability to engineer the real-space potential $V(x,t)$ with modulated amplitudes and phases in time. In experiments with cold atoms, the building block cosine lattice can be formed by laser beams intersecting at an angle \cite{Moritz2003prl,Hadzibabic2004prl,Guo2022prb}.  In experiments with superconducting circuits \cite{Chen2014prb,Hofheinz2011prl,Chen2011apl}, our model can be realized by a microwave cavity in series with a Josephson junction (JJ) biased by a dc voltage ($V$). In this case, the cavity dynamics is  described by the Hamiltonian 
$$\hat{\mathcal{H}}(t)=\hbar\omega_0\hat{a}^\dagger\hat{a}-E_J\cos[\omega_Jt+\Delta(\hat{a}^\dagger+\hat{a})],$$
 where $E_J$ is the JJ energy, $\omega_J=2eV/\hbar$ is the Josephson frequency and $\Delta=\sqrt{2e^2/(\hbar \omega_0C)}$ with $C$ the cavity capacitance \cite{Armour2013prl,Gramich2013prl,Juha2013prl,Juha2015prl,Juha2016prb,Armour2015prb,Trif2015prb,Kubala2015iop,Hofer2016prb,Dambach2017njp}. 
It is a well-established technology in circuit-QED architectures to coherently control multiple tunable Josephson Junctions (JJs) for designing functional quantum devices and quantum computation/simulation, e.g., the Josephson ring modulator architecture \cite{Bergeal2010np,roch2012prl} with $4$ JJs (one for each transmon qubit), the quantum-state-preservation superconducting circuit \cite{Kelly2015nature} with $9$ transmons, the Google programmable superconducting processor Sycamore \cite{Arute2019nat} with $54$ transmon qubits and the recent IBM quantum processor Eagle \cite{Kim2023nat} with $127$ transmons qubit.

 In both experiments, there exists another possible error from implementing the potential by a finite number of laser beams for cold atoms \cite{Guo2022prb} or Josephson junctions \cite{xanda2023arxiv} for superconducting circuits. In our previous work~\cite{guo2024prl}, we have investigated such errors by replacing the integral of wave number in Eq.~(\ref{eq-Vxt-2}) with the sum of a finite number of cosine lattice potentials. The results showed that, although the discretization of the wavenumbers causes some discrepancies during the initial phase of the preparation and also small oscillations in the long-time behavior, the final fidelity of the prepared state keeps high ($>99\%$) even the number of cosine potentials is reduced from one hundred to five.
Note that, our driving scheme could even be realized with a single transmon by decomposing the multiple JJs unitary operation into a sequence of discrete gate operations in the spirit of Trotter discretization \cite{seth1996sci}. A detailed study of this scenario will be a future work {\cite{guo2024arxiv}}.

\subsection{Possible extension  to other Floquet systems}
Although our perturbative framework in this work is tailored for a single driven oscillator, it is possible to extend the present theory to a many-body scenario by upgrading the single-particle plane-wave operator $\exp[{i(k_x\hat{x}+k_p\hat{p})}]$ used in  Eq.~(\ref{eq-HTxp}) to a many-body equivalent $\exp[{\sum_ji(k^j_x\hat{x}_j+k^j_p\hat{p}_j})]$.  In experiments with superconducting circuits, this could be implemented by coupling a dc-voltage biased JJ  to multiple superconducting cavities \cite{Armour2013prl,Armour2015prb,Trif2015prb,Hofer2016prb,Dambach2017njp}.
{ The validity of our NcFT technique for single bosonic mode relies on the one-to-one correspondence between the periodic time and the phase degree of bosonic mode. The problem of extending NcFT from single mode to multiple bosonic modes comes from the fact that there are more than one degrees of phase but only one single time parameter. We have solved this problem by adding proper constraints to the phases and will clarify it in a separate work.  }

Furthermore, we expect our method for bosonic systems can be extended to other Floquet systems that involve spins or fermions. The general idea is to engineer an arbitrary target Hamiltonian in the leading Floquet-Magnus expansion with real experimental conditions. Then, by repeating such technique to high-order Floqut-Magnus expansions, a perturbative framework similar to that shown in Fig.~\ref{Figures-CFE} could be constructed for designing additional high-order correction drives. 

In fact,  Ribeiro et al. \cite{hugo2017prx,Roque2021njp} have developed an alternative perturbative framework for constructing control fields that makes the time evolution at the final moment matches a desired unitary operator with experimental constraints. Relying on the Magnus expansion and the finite Fourier series decomposition of control fields, the problem is reduced to solving a set of linear equations of the Fourier coefficients up to the desired order. As a comparison, our method makes the engineered dynamics matches a desired unitary operator during all the evolution time and provides an iterative framework to calculate the correction drives order by order analytically.

\subsection{Chaos control in classical systems} \label{sec-cccs}
Another interesting prospect of our method is to control chaotic motions in classical systems. Our method formulated in this work is directly for quantum systems. In fact,  our perturbative framework is also valid for classical systems by replacing all the commutators by the Poisson bracket, i.e.,  $[\bullet,\bullet]/(i\lambda)\rightarrow  \{\bullet,\bullet\}$. 
For a generic Hamiltonian system, the chaos comes from the breaking of the regular motions (integrable tori) of system under the perturbation that is resonant with the tori. According to Poincar\'e-Birkhoff theorem~\cite{birkhoff1913pop}, the resonant tori are destroyed by arbitrary small perturbation and split into equal numbers of stable and unstable points.
The non-resonant irrational tori can exist under sufficiently small perturbations but will eventually lose their stability according to Kolmogorov-Arnold-Moser (KAM) theory~\cite{arnold1989book}. 
The instability of the rational tori and KAM tori origins from the resonance among different real modes in physics. 

As exampled by Fig.~\ref{Fig-Hamiltonian}(a), the contour lines of the Hamiltonian Q function represent the oscillator's regular trajectories with some frequency in the classical limit. They will be deformed as the driving strength $\beta$ in Eq.~(\ref{eq-Hcos}) increases. When the deformed counter lines resonate with the high-order Flqouet-Magnus terms, the regular motions will split into several high-order invariant curves (KAM tori), and chaotic regions are separated by KAM tori. 
It was investigated and conjectured that the chaotic bebavior (ergodicity) could be related to the divergence of FM expansion \cite{feldman1984pla,Blanes2009PR,casas2007jpa}, and the radius of convergence for the FM expansion vanishes in the thermodynamic limit \cite{luca2013aop,luca2014prx,moessner2017np}.
From this point of view, the chaotic motion of the system can be suppressed by introducing additional driving potentials that mitigate the higher-order Floquet-Magnus terms. With this control strategy, the regular motions are protected and can survive under a stronger driving strength.

\section{Summary }\label{sec-sum}
In summary, we have developed a general perturbative framework to engineer an arbitrary target Hamiltonian in the Floquet phase space of a periodically driven oscillator beyond RWA. The high-order Floquet-Magnus expansion terms in the engineered Floquet Hamiltonian are mitigated by a systematic perturbative procedure.
Especially, in order to circumvent the problem of calculating the NcFT coefficient of complicated commutators involved in the higher-order Floquet-Magnus terms, we introduced a nontrivial transformation that makes the calculation of high-order corrections feasible. 

We applied our method to a concrete model of a monochromatically driven oscillator for engineering a target Hamiltonian with discrete rotational symmetry and chiral symmetry in phase space. The analytical expression for the 1st-order correction driving potentials is calculated and verified numerically from the engineered quasienery spectrum and eigenstates. The present work aims to establish the general perturbative framework to mitigate errors from higher-order Floquet-Magnus terms. A more technical calculation for the additional driving potentials higher than 1st-order correction, e.g., the 2nd-order Floquet-Magnus expansion given by Eq.~(\ref{eq-HF2}), will be the future work.


\bigskip

{\it Data availability statement.} All data that support the findings of this study are included within the article (and any supplementary files).

{\it Acknowledgments. }  We acknowledge helpful discussions with Vittorio Peano and Florian Marquardt. This work was supported by the National Natural Science Foundation of China (Grant No. 12475025).





\onecolumngrid

\appendix


\section{Q-function of target Hamiltonian  }\label{app-QH}

To calculate the Q-function of  target Hamiltonian $\hat{H}^{(0)}_F$ given by Eq.~(\ref{eq-HF0A2}) in the main text, we first introduce an identity \cite{Liang2018njp} for a monochromatic operator $\hat{M}=\exp[i(s\hat{x}+t\hat{p})]$ with commutative relationship $[\hat{x},\hat{p}]=i\lambda$,
\bea\label{eq-st}
\langle \alpha|\exp\big[i(s\hat{x}+t\hat{p})\big]|\alpha\rangle=\exp\big(-\frac{\lambda}{4}|t-is|^2\big)\exp\big[i(sx+tp)\big],
\eea
where the coordinator and momentum are related to coherent number by
\bea
x\equiv\langle \alpha|\hat{x}|\alpha\rangle=\sqrt{\frac{\lambda}{2}}(\alpha^*+\alpha),\ \ \ \ p\equiv\langle \alpha|\hat{p}|\alpha\rangle=i\sqrt{\frac{\lambda}{2}}(\alpha^*-\alpha).
\eea
Note that the target Hamiltonian (\ref{eq-HF0A2})  is the RWA part of the original Hamiltonian (\ref{eq-Hcos}) in the rotating frame (with frequency $\Omega$) that is given by, cf. Eq.~(\ref{eq-Ht}),
\bea\label{}
\hat{H}(t)= \beta\cos\Big[\hat{x}\cos (\Omega t)+\hat{p}\sin (\Omega t)+n\Omega t\Big]=\frac{\beta}{2}e^{in\Omega t}\exp\Big[\hat{x}\cos (\Omega t)+\hat{p}\sin (\Omega t)\Big]+h.c..
\eea
Using the identity (\ref{eq-st}), we have the Q-function of Hamiltonian $\hat{H}(t)$ as follows
\bea\label{eq-QHt}
\langle \alpha|\hat{H}(t)|\alpha\rangle&=&\frac{\beta}{2}e^{-\frac{\lambda}{4}}e^{in\Omega t}\exp\big(i[x\cos(\Omega t)+p\sin(\Omega t)]\big)+h.c.
=\frac{\beta}{2}e^{-\frac{\lambda}{4}}e^{in\Omega t}e^{ir\cos(\theta-\Omega t)}+h.c.
\eea
Here, in the second line, we have the parameters $(r,\theta)$ via $x=r\cos\theta$ and $p=r\sin\theta$. With the help of well-known Jacobi-Anger expansion $e^{iz\cos\theta}=\sum_{n=-\infty}^{n=+\infty}i^nJ_n(z)e^{in\theta}$ and keeping only the static RWA terms, we have the Q-function of the target Hamiltonian $\hat{H}^{(0)}_F$,
\bea
\langle \alpha|\hat{H}^{(0)}_F|\alpha\rangle=\beta e^{-\frac{\lambda}{4}}J_n(r)\cos(n\theta+\frac{n\pi}{2}).
\eea 

\section{Second-order Floquet-Magnus expansion}\label{app-FM}

The first term $\hat{H}_F^{(2)}(\hat{x},\hat{p})$ on the right-hand side of Eq.~(\ref{eq-HF2}) in the main text is the standard 2nd-order Flqouet-Magnus expansion term \cite{Mikami2016prb}  from the leading-order driving potential $V^{(0)}(x,t)$. In our case, the explicit expression is given by
\bea\label{eq-H2FxpV0}
H^{(2)}_F(\hat{x},\hat{p})&=&\frac{1}{\lambda^2\Omega^2}\sum_{l\neq0}\frac{[\hat{V}^{(0)}_{-l},[\hat{V}^{(0)}_0,\hat{V}^{(0)}_{l}]]}{2l^2}+\frac{1}{\lambda^2\Omega^2}\sum_{l\neq0}\sum_{l'\neq 0,l}\frac{[\hat{V}^{(0)}_{-l'},[\hat{V}^{(0)}_{l'-l},\hat{V}^{(0)}_{l}]]}{3ll'}-\frac{1}{\lambda^2\Omega^2}\sum_{l\neq0}\frac{[\hat{V}^{(0)}_{0},[\hat{V}^{(0)}_0,\hat{V}^{(0)}_{-l}]]}{l^2}e^{il\Omega t_0}\nl
&&-\frac{1}{\lambda^2\Omega^2}\sum_{l,l'\neq0}\frac{[\hat{V}^{(0)}_{l'},[\hat{V}^{(0)}_{-l'},\hat{V}^{(0)}_{-l}]]}{3ll'}e^{il\Omega t_0}+\frac{1}{\lambda^2\Omega^2}\sum_{l,l'\neq0}\frac{[\hat{V}^{(0)}_{-l},[\hat{V}^{(0)}_{l'},\hat{V}^{(0)}_{-l'}]]}{3ll'}e^{il\Omega t_0}\nl
&&-\frac{1}{\lambda^2\Omega^2}\sum_{l\neq0}\sum_{l'\neq 0,l}\frac{[\hat{V}^{(0)}_{0},[\hat{V}^{(0)}_{l'-l},\hat{V}^{(0)}_{-l'}]]}{2ll'}e^{il\Omega t_0}+\frac{1}{\lambda^2\Omega^2}\sum_{l,l'\neq0}\frac{[\hat{V}^{(0)}_{0},[\hat{V}^{(0)}_{-l'},\hat{V}^{(0)}_{-l}]]}{2ll'}e^{i(l+l')\Omega t_0}\nl
&&-\frac{1}{\lambda^2\Omega^2}\sum_{l,l'\neq0}\frac{[\hat{V}^{(0)}_{-l'},[\hat{V}^{(0)}_{0},\hat{V}^{(0)}_{-l}]]}{2ll'}e^{i(l+l')\Omega t_0}.
 \eea

\section{NcFT coefficient of commutators}\label{app-hom}

We present detailed derivation for the transformation given by Eqs.~(\ref{eq-main-1})-(\ref{eq-fijk}) in the main text that can circumvent the difficulty to calculate the commutators of  harmonics in the higher-order Floquet-Magnus expansions and directly calculate the NcFT coefficient of commutators. 

\subsection{General form}\label{app-general}

We can write any time-periodic Hamiltonian in the NcFT formula as follows 
\bea
H(t)&=& \int_{-\infty}^{+\infty}\frac{|k|}{2}f(k,\Omega t)e^{ik[\hat{P}\sin (\Omega t)+\hat{X}\cos (\Omega t)]}dk.
\eea
The harmonics $H_l$ defined via
$
H(t)=\sum_{l\in \mathbb{Z}}H_le^{il\Omega t}
$
can be calculated by
\bea
H_{l'}(\hat{X},\hat{P})&=&\frac{1}{T}\int_0^Tdt' \int_{-\infty}^{+\infty}dk'\frac{|k'|}{2}f(k',\Omega t')e^{-i\Omega l' t'}e^{ik'[\hat{P}\sin (\Omega t')+\hat{X}\cos (\Omega t')]}\nl
H_{l''}(\hat{X},\hat{P})&=&\frac{1}{T}\int_0^Tdt'' \int_{-\infty}^{+\infty}dk''\frac{|k''|}{2}f(k'',\Omega t'')e^{-i\Omega l''t''}e^{ik''[\hat{P}\sin (\Omega t'')+\hat{X}\cos (\Omega t'')]}\nl
H_{l'}H_{l''}&=&\frac{1}{T^2}\int_0^T dt''\int_0^T dt'\int_{-\infty}^{+\infty}dk'' \int_{-\infty}^{+\infty}dk'\frac{|k'k''|}{4}f(k',\Omega t')f(k'',\Omega t'')e^{-i\Omega (l' t'+l''t'')}\nl
&&\times\exp\big(ik'[\hat{P}\sin (\Omega t')+\hat{X}\cos (\Omega t')]\big)\exp\big(ik''[\hat{P}\sin (\Omega t'')+\hat{X}\cos (\Omega t'')]\big)\nl
&=&\frac{1}{T^2}\int_0^T dt''\int_0^T dt'\int_{-\infty}^{+\infty}dk'' \int_{-\infty}^{+\infty}dk'\frac{|k'k''|}{4}f(k',\Omega t')f(k'',\Omega t'')e^{-i\Omega (l' t'+l''t'')}\nl
&&\times\exp\big(i\hat{P}[k''\sin (\Omega t'')+k'\sin (\Omega t')]+i\hat{X}[k''\cos (\Omega t'')+k'\cos (\Omega t')]\big){e^{i\lambda\frac{k''k'}{2}\sin\Omega(t'-t'')}}\nl
H_{l''}H_{l'}&=&\frac{1}{T^2}\int_0^T dt''\int_0^T dt'\int_{-\infty}^{+\infty}dk'' \int_{-\infty}^{+\infty}dk'\frac{|k'k''|}{4}f(k',\Omega t')f(k'',\Omega t'')e^{-i\Omega (l' t'+l''t'')}\nl
&&\times\exp\big(ik''[\hat{P}\sin (\Omega t'')+\hat{X}\cos (\Omega t'')]\big)\exp\big(ik'[\hat{P}\sin (\Omega t')+\hat{X}\cos (\Omega t')]\big)\nl
&=&\frac{1}{T^2}\int_0^T dt''\int_0^T dt'\int_{-\infty}^{+\infty}dk'' \int_{-\infty}^{+\infty}dk'\frac{|k'k''|}{4}f(k',\Omega t')f(k'',\Omega t'')e^{-i\Omega (l' t'+l''t'')}\nl
&&\times\exp\big(i\hat{P}[k''\sin (\Omega t'')+k'\sin (\Omega t')]+i\hat{X}[k''\cos (\Omega t'')+k'\cos (\Omega t')]\big){e^{-i\lambda\frac{k''k'}{2}\sin\Omega(t'-t'')}}.
\eea
Therefore, we have
\bea
H_{l'}H_{l''}-H_{l''}H_{l'}&=&\frac{i}{T^2}\int_0^T dt''\int_0^T dt'\int_{-\infty}^{+\infty}dk'' \int_{-\infty}^{+\infty}dk'\frac{|k'k''|}{2}f(k',\Omega t')f(k'',\Omega t'')e^{-i\Omega (l' t'+l''t'')}\nl
&&\times\exp\Big(i\hat{X}[k''\cos (\Omega t'')+k'\cos (\Omega t')]+i\hat{P}[k''\sin (\Omega t'')+k'\sin (\Omega t')]\Big)\nl
&&\times {\sin\Big[\lambda\frac{k''k'}{2}\sin\Omega(t'-t'')\Big]}.
\eea
By introducing new variables 
\bea
\left\{
\begin{array}{lll}
k_1&=&k'\cos (\Omega t')+k''\cos (\Omega t'')\\
k_2&=&k'\sin (\Omega t')+k''\sin (\Omega t'')\\
dk_1dk_2&=&{\big|\sin[\Omega(t'-t'')]\big|}dk''dk',
\end{array}\right.
\eea
and the inverse transformation
\bea\label{eq-k''k'k}
\left\{
\begin{array}{lll}
k'&=&\frac{k_1\sin (\Omega t'')-k_2\cos (\Omega t'')}{\sin[\Omega(t''-t')]}=k\frac{\sin(\Omega t''-\theta)}{\sin[\Omega(t''-t')]},\\
k''&=&\frac{k_1\sin (\Omega t')-k_2\cos (\Omega t')}{\sin[\Omega(t'-t'')]}=k\frac{\sin(\Omega t'-\theta)}{\sin[\Omega(t'-t'')]}
\end{array}\right.
\eea
with ($k_1=k\cos\theta$, $k_2=k\sin\theta$), we have
\bea\label{eq-Hl'Hl''}
H_{l'}H_{l''}-H_{l''}H_{l'}&=&\frac{1}{2\pi}\int_{-\infty}^{+\infty}dk_1 \int_{-\infty}^{+\infty}dk_2\exp\big(ik_1\hat{X}+ik_2\hat{P}\big)\nl
&\times&\Bigg(\frac{2\pi i}{T^2}\int_0^T dt''\int_0^T dt'\frac{|k'k''|}{2}f(k',\Omega t')f(k'',\Omega t'')e^{-i\Omega (l' t'+l''t'')}\frac{\sin\big[\lambda\frac{k''k'}{2}\sin\Omega(t'-t'')\big]}{{\big|\sin[\Omega(t'-t'')]\big|}}\Bigg).\ \ \ \ 
\eea
The above Eqs.~(\ref{eq-k''k'k})-(\ref{eq-Hl'Hl''}) are the transformation given by Eqs.~(\ref{eq-main-1})-(\ref{eq-fijk}) in the main text.

\subsection{Jacobian matrix}
For further discussion below, we calculate the Jacobian matrix for fixed $k_1$ and $k_2$ as follows
\bea
J&=&\frac{\partial(k',k'')}{\partial(t',t'')}\equiv
\begin{pmatrix}
\begin{array}{cc}
\partial_{t'}k' & \partial_{t''}k'  \\
\partial_{t'}k'' & \partial_{t''}k''
\end{array} 
\end{pmatrix}
\nl
&=&
\begin{pmatrix}
k\Omega\frac{\sin(\Omega t''-\theta)\cos[\Omega(t''-t')]}{\sin^2[\Omega(t''-t')]} &k\Omega\frac{\cos(\Omega t''-\theta)\sin[\Omega(t''-t')]-\sin(\Omega t''-\theta)\cos[\Omega(t''-t')]}{\sin^2[\Omega(t''-t')]}  \\
k\Omega\frac{\cos(\Omega t'-\theta)\sin[\Omega(t'-t'')]-\sin(\Omega t'-\theta)\cos[\Omega(t'-t'')]}{\sin^2[\Omega(t'-t'')]}  & k\Omega\frac{\sin(\Omega t'-\theta)\cos[\Omega(t'-t'')]}{\sin^2[\Omega(t'-t'')]}
\end{pmatrix}
\nl
&=&
\left( 
\begin{array}{cc}
k\Omega\frac{\sin(\Omega t''-\theta)\cos[\Omega(t''-t')]}{\sin^2[\Omega(t''-t')]} &k\Omega\frac{-\sin[\Omega t'-\theta]}{\sin^2[\Omega(t''-t')]}  \\
k\Omega\frac{-\sin[\Omega t''-\theta]}{\sin^2[\Omega(t'-t'')]}  & k\Omega\frac{\sin(\Omega t'-\theta)\cos[\Omega(t'-t'')]}{\sin^2[\Omega(t'-t'')]}
\end{array} 
\right)
\eea
and the Jacobian determinant is
\bea\label{eq-detJ}
\mathrm{det}(J)&=&(k\Omega)^2\frac{\sin(\Omega t''-\theta)\sin(\Omega t'-\theta)}{\sin^4[\Omega(t'-t'')]}\Big(\cos^2[\Omega(t'-t'')]-1\Big)
=-(k\Omega)^2\frac{\sin(\Omega t''-\theta)\sin(\Omega t'-\theta)}{\sin^4[\Omega(t'-t'')]}\sin^2[\Omega(t'-t'')]\nl
&=&-(k\Omega)^2\frac{\sin(\Omega t''-\theta)\sin(\Omega t'-\theta)}{\sin^2[\Omega(t'-t'')]}.
\eea
When $t'= \theta/\Omega\pm 1/\Omega\arccos[k/2], \ t''= \theta/\Omega\mp 1/\Omega\arccos[k/2]$, we have $|\mathrm{det}(J)|=\Omega^2$.

\subsection{1st-order Floquet-Magnus expansion}

We write the $l$-th term in the first-order Magnus expansion given by Eq.~(\ref{eq-H1Fxp}) (by taking $l'=-l''=l$)
\bea\label{eq-1stME}
\frac{1}{\lambda\Omega l}[H_l,H_{-l}]&\equiv&\frac{1}{2\pi}\int_{-\infty}^{+\infty}dk_1 \int_{-\infty}^{+\infty}dk_2f_{l,-l}(k_1,k_2)e^{ik_1\hat{X}+ik_2\hat{P}}
\eea
with
\bea\label{eq-fl-l}
f_{l,-l}(k_1,k_2)&=&f_{l,-l}(k,\theta)\nl
&\equiv&\frac{2\pi i}{\lambda\Omega l T^2}\int_0^T dt''\int_0^T dt'\frac{|k'k''|}{2}f(k',\Omega t')f(k'',\Omega t'')e^{-i\Omega l( t'-t'')}\frac{\sin\big[\lambda\frac{k''k'}{2}\sin\Omega(t'-t'')\big]}{ {\big|\sin[\Omega(t'-t'')]\big|}}.
\eea
One can prove the following properties by exchanging $t'$ and $t''$ and using $f^*(k,\Omega t)=f(-k,\Omega t)$,
\bea
f_{l,-l}(k_1,k_2)&=&f_{-l,l}(k_1,k_2)\nl
f^*_{l,-l}(k_1,k_2)&=&f_{l,-l}(-k_1,-k_2).
\eea
{
Another $l$-th term in Eq.~(\ref{eq-H1Fxp}) is 
\bea\label{eq-1stME-2}
\frac{1}{\lambda\Omega l}[\hat{H}_{-l},\hat{H}_{0}]e^{il\Omega t_0}&\equiv&\frac{1}{2\pi}\int_{-\infty}^{+\infty}dk_1 \int_{-\infty}^{+\infty}dk_2f_{-l,0}(k_1,k_2)e^{ik_1\hat{X}+ik_2\hat{P}}
\eea
with
\bea\label{eq-fl-l-2}
f_{-l,0}(k_1,k_2)&=&f_{-l,0}(k,\theta)\nl
&\equiv&\frac{2\pi i}{\lambda\Omega l T^2}\int_0^T dt''\int_0^T dt'\frac{|k'k''|}{2}f(k',\Omega t')f(k'',\Omega t'')e^{i\Omega l (t'+t_0)}\frac{\sin\big[\lambda\frac{k''k'}{2}\sin\Omega(t'-t'')\big]}{ {\big|\sin[\Omega(t'-t'')]\big|}}.
\eea
By taking $l'=l,l''=0$, we have another term in Eq.~(\ref{eq-H1Fxp})
\bea\label{eq-1stME-20}
-\frac{1}{\lambda\Omega l}[\hat{H}_{l},\hat{H}_{0}]e^{-il\Omega t_0}&=&\Big(\frac{1}{\lambda\Omega l}[\hat{H}_{-l},\hat{H}_{0}]e^{il\Omega t_0}\Big)^\dagger
\equiv\frac{1}{2\pi}\int_{-\infty}^{+\infty}dk_1 \int_{-\infty}^{+\infty}dk_2f^*_{-l,0}(-k_1,-k_2)e^{ik_1\hat{X}+ik_2\hat{P}}.
\eea
}

\section{Monochromatically driven harmonic oscillator}\label{app-potential}
The Hamiltonian of a monochromatically driven harmonic oscillator is given by
\bea\label{eq-tildeH}
\tilde{H}(t)=\frac{1}{2}(\hat{x}^2+\hat{p}^2)+A\cos(\hat{x}+n\Omega t).
\eea
According to Eq.~(\ref{eq-Vxt-1}), we have
\bea
f(k,\Omega t)=A\delta(k-1)e^{in\Omega t}+A\delta(k+1)e^{-in\Omega t}.
\eea
To proceed, we introduce some properties of Dirac functions. The composition $\delta(g(x))$ for continuously differentiable functions $g(x)$ is defined by
\bea
\delta(g(x)) = \sum_i \frac{\delta(x-x_i)}{|dg(x_i)/dx|}
\eea
where the sum extends over all roots (i.e., all the different ones) of $g(x)$, which are assumed to be simple root simple. For multiple component function, the Dirac funtion is
\bea\label{eq-deltas}
\delta(g(x,y))\delta(h(x,y))=\sum_{i}\frac{\delta(x-x_i)\delta(y-y_i)}{|\partial_x g(x_i,y_i)\partial _yh(x_i,y_i)-\partial_yg(x_i,y_i)\partial_xh(x_i,y_i)|},
\eea
where $x_i$, $y_i$ are the roots that satisfying $g(x_i,y_i)=0$ and $h(x_i,y_i)=0$.

\subsection{Calculation of $f_{l,-l}(k,\theta)$}\label{app-potential-fll}
Plugging the above expression into Eq.~(\ref{eq-fl-l}), we have
\bea\label{eq-fl-lk}
f_{l,-l}(k,\theta)&=&\frac{2\pi i}{\lambda\Omega lT^2}\int_0^T dt''\int_0^T dt'\frac{|k'k''|}{2}f(k',\Omega t')f(k'',\Omega t'')e^{-il\Omega ( t'-t'')}\frac{\sin\big[\lambda\frac{k''k'}{2}\sin\Omega(t'-t'')\big]}{|\sin[\Omega(t'-t'')]|}\\
&=&\frac{2\pi iA^2}{2\lambda\Omega lT^2}\int_0^T dt''\int_0^T dt'|k'k''|\Big[\delta(k'-1)\delta(k''-1)e^{in\Omega (t'+t'')}+\delta(k'-1)\delta(k''+1)e^{in\Omega (t'-t'')}\nl
&&+\delta(k'+1)\delta(k''-1)e^{-in\Omega (t'-t'')}+\delta(k'+1)\delta(k''+1)e^{-in\Omega (t'+t'')}\Big]
e^{-i\Omega l( t'-t'')}\frac{\sin\big[\lambda\frac{k''k'}{2}\sin\Omega(t'-t'')\big]}{|\sin[\Omega(t'-t'')]|}.\nn
\eea
According to Eqs.~(\ref{eq-k''k'k}) and (\ref{eq-deltas}), we have
\bea\label{eq-deltak'}
\delta(k'(t',t'')\pm 1)\delta(k''(t',t'')\pm 1)=\sum_{i}\frac{\delta(t'-t'_i)\delta(t''-t''_i)}{|\partial_{t'} k'\partial _{t''}k''-\partial_{t''}k'\partial_{t'}k''|_{(t'_i,t''_i)}}=\sum_{i}\frac{\delta(t'-t'_i)\delta(t''-t''_i)}{|\mathrm{det}(J)|_{(t'_i,t''_i)}}.
\eea
where $t'_i$, $t''_i$ are the roots that satisfying $k'(t'_i,t''_i)=\mp 1$ and $k''(t'_i,t''_i)=\mp 1$.  According to Eq.~(\ref{eq-k''k'k}), we have the following solutions for the given value of $k$ and $\theta$.

(1) For the case of $k'=k''=1$, we have from Eq.~(\ref{eq-k''k'k})
\begin{eqnarray}\label{eq-t't''k}
\frac{\sin(\Omega t''-\theta)}{\sin[\Omega(t''-t')]}=\frac{1}{k},\ \ \ 
\frac{\sin(\Omega t'-\theta)}{\sin[\Omega(t'-t'')]}=\frac{1}{k}.
\end{eqnarray}
Comparing the above two equations, we have $\sin(\Omega t''-\theta)=-\sin(\Omega t'-\theta)$. By assuming $t'=\theta/\Omega+\alpha/\Omega$ and $t''=\theta/\Omega-\alpha/\Omega$, we have $\cos\alpha=k/2$.
The roots are 
\bea
t'_i=\frac{\theta}{\Omega}\pm\frac{1}{\Omega}\arccos\Big(\frac{k}{2}\Big),\ \ \  t''_i=\frac{\theta}{\Omega}\mp\frac{1}{\Omega}\arccos\Big(\frac{k}{2}\Big).
\eea
One may also wonder another type solution of $t''=t'+\pi/\Omega$ which also satisfies $\sin(\Omega t''-\theta)=-\sin(\Omega t'-\theta)$. However, there is no such kind of solution for a nonzero $k$ even in the limit sense. For example, we assume $t'=\theta/\Omega+\epsilon'/\Omega$ and $t''=t'+\pi/\Omega+\epsilon''/\Omega$ where $\epsilon',\epsilon''\rightarrow0$. Plugging them back to Eq.~(\ref{eq-t't''k}), we have 
$
\frac{\epsilon'+\epsilon''}{\epsilon''}=\frac{1}{k},\  \frac{\epsilon'}{\epsilon''}=\frac{1}{k}.
$
But these two conditions are obviously contradictory with each other. The same argument also applies for the case of $t'=\theta/\Omega+\pi/\Omega+\epsilon'/\Omega$.

(2) For the case of $k'=k''=-1$, the condition Eq.~(\ref{eq-t't''k}) becomes \begin{eqnarray}\label{}
\frac{\sin(\Omega t''-\theta)}{\sin[\Omega(t''-t')]}=-\frac{1}{k},\ \ \ 
\frac{\sin(\Omega t'-\theta)}{\sin[\Omega(t'-t'')]}=-\frac{1}{k}.
\end{eqnarray}
The roots are 
\bea
t'_i=\frac{\theta}{\Omega}\pm\frac{1}{\Omega}\arccos\Big(-\frac{k}{2}\Big),\ \ \  t''_i=\frac{\theta}{\Omega}\mp\frac{1}{\Omega}\arccos\Big(-\frac{k}{2}\Big).
\eea

(3) For the case of $k'=1,k''=-1$, 
the condition Eq.~(\ref{eq-t't''k}) becomes \begin{eqnarray}\label{}
\frac{\sin(\Omega t''-\theta)}{\sin[\Omega(t''-t')]}=\frac{1}{k},\ \ \ 
\frac{\sin(\Omega t'-\theta)}{\sin[\Omega(t'-t'')]}=-\frac{1}{k}.
\end{eqnarray}
We have the condition that $\sin(\Omega t''-\theta)=\sin(\Omega t'-\theta)$. For the case of $t'=t''$, there are no such kind of roots. For the case of $\Omega t''-\theta=\pi-(\Omega t'-\theta)$, i.e., $t''=-t'+(2\theta+\pi)/\Omega$, the condition becomes $\cos(\Omega t'-\theta)=k/2$ and the roots are
The roots are 
\bea
t'_i=\frac{\theta}{\Omega}\pm\frac{1}{\Omega}\arccos\Big(\frac{k}{2}\Big),\ \ \  t''_i=\frac{\theta+\pi}{\Omega}\mp\frac{1}{\Omega}\arccos\Big(\frac{k}{2}\Big).
\eea

(4) For the case of $k'=-1,k''=1$, 
the condition Eq.~(\ref{eq-t't''k}) becomes \begin{eqnarray}\label{}
\frac{\sin(\Omega t''-\theta)}{\sin[\Omega(t''-t')]}=-\frac{1}{k},\ \ \ 
\frac{\sin(\Omega t'-\theta)}{\sin[\Omega(t'-t'')]}=\frac{1}{k}.
\end{eqnarray}
We have the condition that $\sin(\Omega t''-\theta)=\sin(\Omega t'-\theta)$. For the case of $t'=t''$, there are no such kind of roots. For the case of $\Omega t''-\theta=\pi-(\Omega t'-\theta)$, i.e., $t''=-t'+(2\theta+\pi)/\Omega$, the condition becomes $\cos(\omega t'-\theta)=-k/2$ and the roots are
The roots are 
\bea
t'_i=\frac{\theta}{\Omega}\pm\frac{1}{\Omega}\arccos\Big(-\frac{k}{2}\Big),\ \ \  t''_i=\frac{\theta+\pi}{\Omega}\mp\frac{1}{\Omega}\arccos\Big(-\frac{k}{2}\Big).
\eea

From Eqs.~(\ref{eq-detJ}) and (\ref{eq-deltak'}), the expression Eq.~(\ref{eq-fl-lk}) is
\bea
&&f_{l,-l}(k,\theta)\nl
&=&\frac{2\pi i}{\lambda\Omega lT^2}\int_0^T dt''\int_0^T dt'\frac{|k'k''|}{2}f(k',\Omega t')f(k'',\Omega t'')e^{-il\Omega ( t'-t'')}\frac{\sin\big[\lambda\frac{k''k'}{2}\sin\Omega(t'-t'')\big]}{|\sin[\Omega(t'-t'')]|}\nl
&=&\frac{2\pi iA^2}{2\lambda\Omega lT^2}\int_0^T dt''\int_0^T dt'|k'k''|e^{-i\Omega l( t'-t'')}\frac{\sin\big[\lambda\frac{k''k'}{2}\sin\Omega(t'-t'')\big]}{|\sin[\Omega(t'-t'')]|}\nl
&&\Big[\delta(k'-1)\delta(k''-1)e^{in\Omega (t'+t'')}+\delta(k'-1)\delta(k''+1)e^{in\Omega (t'-t'')}\nl
&&+\delta(k'+1)\delta(k''-1)e^{-in\Omega (t'-t'')}+\delta(k'+1)\delta(k''+1)e^{-in\Omega (t'+t'')}\Big]
\nl
&=&\frac{2\pi iA^2}{2\lambda\Omega lT^2}\sum_{i,j}\int_0^T dt''\int_0^T dt'\delta(t'-t'_i)\delta(t''-t''_j)\nl
&&e^{-i\Omega l( t'_i-t''_j)}\frac{\sin\big[\lambda\frac{k''k'}{2}\sin\Omega(t'_i-t''_j)\big]}{|\sin[\Omega(t'_i-t''_j)]|}\frac{|k'k''|}{|\mathrm{det}(J)|_{(t'_i,t''_i)}}\nl
&&\Big[e^{in\Omega (t'_i+t''_j)}|_{k'=1,k''=1}+e^{-in\Omega (t'_i+t''_j)}|_{k'=-1,k''=-1}\nl
&&+e^{in\Omega (t'_i-t''_j)}|_{k'=1,k''=-1}+e^{-in\Omega (t'_i-t''_j)}|_{k'=-1,k''=1}\Big]
\nl
&=&\frac{2\pi iA^2}{ T^2}\frac{1}{2\lambda\Omega l}\nl
&&\Bigg(e^{i2n\theta-i2l\arccos\frac{k}{2}}\frac{\sin [\frac{\lambda}{2}\sin(2\arccos\frac{k}{2})]}{|\sin(2\arccos\frac{k}{2})|}\frac{1}{\Omega^2}-e^{i2n\theta+i2l\arccos\frac{k}{2}}\frac{\sin [\frac{\lambda}{2}\sin(2\arccos\frac{k}{2})]}{|\sin(2\arccos\frac{k}{2})|}\frac{1}{\Omega^2}\nl
&&+e^{-i2n\theta-i2l\arccos(-\frac{k}{2})}\frac{\sin (\frac{\lambda}{2}\sin[2\arccos(-\frac{k}{2})])}{|\sin[2\arccos(-\frac{k}{2})]|}\frac{1}{\Omega^2}-e^{-i2n\theta+i2l\arccos(-\frac{k}{2})}\frac{\sin (\frac{\lambda}{2}\sin[2\arccos(-\frac{k}{2})])}{|\sin[2\arccos(-\frac{k}{2})]|}\frac{1}{\Omega^2}\Bigg)\nl
&&+e^{i(n-l)(2\arccos\frac{k}{2}-\pi)}\frac{\sin [\frac{\lambda}{2}\sin(2\arccos\frac{k}{2})]}{|\sin(2\arccos\frac{k}{2})|}\frac{1}{\Omega^2}-e^{i(n-l)(-2\arccos\frac{k}{2}-\pi)}\frac{\sin [\frac{\lambda}{2}\sin(2\arccos\frac{k}{2})]}{|\sin(2\arccos\frac{k}{2})|}\frac{1}{\Omega^2}\nl
&&+e^{-i(n+l)[2\arccos(-\frac{k}{2})-\pi]}\frac{\sin (\frac{\lambda}{2}\sin[2\arccos(-\frac{k}{2})])}{|\sin[2\arccos(-\frac{k}{2})]|}\frac{1}{\Omega^2}-e^{-i(n+l)[-2\arccos(-\frac{k}{2})-\pi]}\frac{\sin (\frac{\lambda}{2}\sin[2\arccos(-\frac{k}{2})])}{|\sin[2\arccos(-\frac{k}{2})]|}\frac{1}{\Omega^2}\Bigg)\nl
&=&\frac{2\pi iA^2}{ T^2}\frac{1}{2\lambda\Omega l}\frac{\sin [\frac{\lambda}{2}\sin(2\arccos\frac{k}{2})]}{|\sin(2\arccos\frac{k}{2})|}\frac{1}{\Omega^2}\nl
&&\Big(e^{i2n\theta-i2l\arccos\frac{k}{2}}-e^{i2n\theta+i2l\arccos\frac{k}{2}}-e^{-i2n\theta-i2l\arccos(-\frac{k}{2})}+e^{-i2n\theta+i2l\arccos(-\frac{k}{2})}\nl
&&+e^{i(n-l)[\arccos\frac{k}{2}-\arccos(-\frac{k}{2})]}-e^{-i(n-l)[\arccos\frac{k}{2}-\arccos(-\frac{k}{2})]}\nl
&&-e^{i(n+l)[\arccos\frac{k}{2}-\arccos(-\frac{k}{2})]}+e^{-i(n+l)[\arccos\frac{k}{2}-\arccos(-\frac{k}{2})]}\Big)\nl
&=&\frac{2\pi iA^2}{ T^2}\frac{1}{2\lambda\Omega l}\frac{\sin [\frac{\lambda}{2}\sin(2\arccos\frac{k}{2})]}{|\sin(2\arccos\frac{k}{2})|}\frac{1}{\Omega^2}\nl
&&\Big[-4i\cos(2n\theta)\sin(2l\arccos\frac{k}{2})-4i(-1)^{n+l}\cos(2n\arccos\frac{k}{2})\sin(2l\arccos\frac{k}{2})\Big]\nl
&=&\frac{2\pi A^2}{ (\Omega T)^2}\frac{2}{\lambda\Omega l}\frac{\sin [\frac{\lambda}{2}\sin(2\arccos\frac{k}{2})]}{|\sin(2\arccos\frac{k}{2})|}\sin(2l\arccos\frac{k}{2})\Big[\cos(2n\theta)+(-1)^{n+l}\cos(2n\arccos\frac{k}{2})\Big].
\eea
It can be seen directly 
\bea
f_{-l,l}(k,\theta)=f_{l,-l}(k,\theta),\ \ \ \ f_{-l,l}(-k,\theta)=f^*_{l,-l}(k,\theta).
\eea

\subsection{Calculation of $f_{-l,0}$}

We present a detailed calculation for $f_{-l,0}$ from the formula given by Eq.~(\ref{eq-fl-l-2}) as follows:
\bea\label{}
&&f_{-l,0}(k,\theta)\nl
&\equiv&\frac{2\pi i}{\lambda\Omega l T^2}\int_0^T dt''\int_0^T dt'\frac{|k'k''|}{2}f(k',\Omega t')f(k'',\Omega t'')e^{i\Omega l (t'+t_0)}\frac{\sin\big[\lambda\frac{k''k'}{2}\sin\Omega(t'-t'')\big]}{ {\big|\sin[\Omega(t'-t'')]\big|}}\nl
&=&\frac{2\pi iA^2}{2\lambda\Omega lT^2}\int_0^T dt''\int_0^T dt'|k'k''|\nl
&&\Big[\delta(k'-1)\delta(k''-1)e^{in\Omega (t'+t'')}+\delta(k'-1)\delta(k''+1)e^{in\Omega (t'-t'')}\nl
&&+\delta(k'+1)\delta(k''-1)e^{-in\Omega (t'-t'')}+\delta(k'+1)\delta(k''+1)e^{-in\Omega (t'+t'')}\Big]
\nl
&&e^{i\Omega l( t'+t_0)}\frac{\sin\big[\lambda\frac{k''k'}{2}\sin\Omega(t'-t'')\big]}{|\sin[\Omega(t'-t'')]|}\nl
&=&\frac{2\pi iA^2}{2\lambda\Omega lT^2}\sum_{i,j}\int_0^T dt''\int_0^T dt'\delta(t'-t'_i)\delta(t''-t''_j)\nl
&&e^{i\Omega l( t'_i+t_0)}\frac{\sin\big[\lambda\frac{k''k'}{2}\sin\Omega(t'_i-t''_j)\big]}{|\sin[\Omega(t'_i-t''_j)]|}\frac{|k'k''|}{|\mathrm{det}(J)|_{(t'_i,t''_i)}}\nl
&&\Big[e^{in\Omega (t'_i+t''_j)}|_{k'=1,k''=1}+e^{-in\Omega (t'_i+t''_j)}|_{k'=-1,k''=-1}+e^{in\Omega (t'_i-t''_j)}|_{k'=1,k''=-1}+e^{-in\Omega (t'_i-t''_j)}|_{k'=-1,k''=1}\Big]
\nl
&=&\frac{2\pi iA^2}{ T^2}\frac{1}{2\lambda\Omega l}\nl
&&\Bigg(e^{i2n\theta+il(\theta+\Omega t_0+\arccos\frac{k}{2})}\frac{\sin [\frac{\lambda}{2}\sin(2\arccos\frac{k}{2})]}{|\sin(2\arccos\frac{k}{2})|}\frac{1}{\Omega^2}-e^{i2n\theta+il(\theta+\Omega t_0-\arccos\frac{k}{2})}\frac{\sin [\frac{\lambda}{2}\sin(2\arccos\frac{k}{2})]}{|\sin(2\arccos\frac{k}{2})|}\frac{1}{\Omega^2}\nl
&&+e^{-i2n\theta+il[\theta+\Omega t_0+\arccos(-\frac{k}{2})]}\frac{\sin (\frac{\lambda}{2}\sin[2\arccos(-\frac{k}{2})])}{|\sin[2\arccos(-\frac{k}{2})]|}\frac{1}{\Omega^2}-e^{-i2n\theta+il[\theta+\Omega t_0-\arccos(-\frac{k}{2})]}\frac{\sin (\frac{\lambda}{2}\sin[2\arccos(-\frac{k}{2})])}{|\sin[2\arccos(-\frac{k}{2})]|}\frac{1}{\Omega^2}\nl
&&+e^{in(2\arccos\frac{k}{2}-\pi)+il(\theta+\Omega t_0+\arccos\frac{k}{2})}\frac{\sin [\frac{\lambda}{2}\sin(2\arccos\frac{k}{2})]}{|\sin(2\arccos\frac{k}{2})|}\frac{1}{\Omega^2}\nl
&&-e^{in(-2\arccos\frac{k}{2}-\pi)+il(\theta+\Omega t_0-\arccos\frac{k}{2})}\frac{\sin [\frac{\lambda}{2}\sin(2\arccos\frac{k}{2})]}{|\sin(2\arccos\frac{k}{2})|}\frac{1}{\Omega^2}\nl
&&+e^{-in[2\arccos(-\frac{k}{2})-\pi]+il[\theta+\Omega t_0+\arccos(-\frac{k}{2})]}\frac{\sin (\frac{\lambda}{2}\sin[2\arccos(-\frac{k}{2})])}{|\sin[2\arccos(-\frac{k}{2})]|}\frac{1}{\Omega^2}\nl
&&-e^{-in[-2\arccos(-\frac{k}{2})-\pi]+il[\theta+\Omega t_0-\arccos(-\frac{k}{2})]}\frac{\sin (\frac{\lambda}{2}\sin[2\arccos(-\frac{k}{2})])}{|\sin[2\arccos(-\frac{k}{2})]|}\frac{1}{\Omega^2}\Bigg)\nl
&=&\frac{2\pi iA^2}{ T^2}\frac{1}{2\lambda\Omega l}\frac{\sin [\frac{\lambda}{2}\sin(2\arccos\frac{k}{2})]}{|\sin(2\arccos\frac{k}{2})|}\frac{1}{\Omega^2}\nl
&&\Big[e^{i2n\theta+il(\theta+\Omega t_0+\arccos\frac{k}{2})}-e^{i2n\theta+il(\theta+\Omega t_0-\arccos\frac{k}{2})}
-e^{-i2n\theta+il[\theta+\Omega t_0+\arccos(-\frac{k}{2})]}+e^{-i2n\theta+il[\theta+\Omega t_0-\arccos(-\frac{k}{2})]}\nl
&&+e^{in(2\arccos\frac{k}{2}-\pi)+il(\theta+\Omega t_0+\arccos\frac{k}{2})}-e^{in(-2\arccos\frac{k}{2}-\pi)+il(\theta+\Omega t_0-\arccos\frac{k}{2})}\nl
&&-e^{-in[2\arccos(-\frac{k}{2})-\pi]+il[\theta+\Omega t_0+\arccos(-\frac{k}{2})]}+e^{-in[-2\arccos(-\frac{k}{2})-\pi]+il[\theta+\Omega t_0-\arccos(-\frac{k}{2})]}\Big]\nl
&=&\frac{2\pi iA^2}{ T^2}\frac{1}{2\lambda\Omega l}\frac{\sin [\frac{\lambda}{2}\sin(2\arccos\frac{k}{2})]}{|\sin(2\arccos\frac{k}{2})|}\frac{1}{\Omega^2}e^{il(\theta+\Omega t_0)}\nl
&&2i\Big[e^{i2n\theta}\sin[l\arccos\frac{k}{2}]
-e^{-i2n\theta}\sin[l\arccos(-\frac{k}{2})]+e^{-in\pi}\sin[(2n+l)\arccos\frac{k}{2}]+e^{in\pi}\sin[(2n-l)\arccos(-\frac{k}{2})]\Big]\nl
&=&-\frac{2\pi A^2}{ (\Omega T)^2}\frac{1}{\lambda\Omega l}\frac{\sin [\frac{\lambda}{2}\sin(2\arccos\frac{k}{2})]}{|\sin(2\arccos\frac{k}{2})|}e^{il\Omega t_0}
\Big[e^{i(2n+l)\theta}\sin[l\arccos\frac{k}{2}]
-e^{-i(2n-l)\theta}\sin[l\arccos(-\frac{k}{2})]\nl
&&+e^{il\theta}\Big(e^{-in\pi}\sin[(2n+l)\arccos\frac{k}{2}]+e^{in\pi}\sin[(2n-l)\arccos(-\frac{k}{2})]\Big)\Big].
\eea

\section{Eigenproblem of Floquet system }\label{app-FS}

We provide numerical details to solve the eigenproblem of a Floquet system whose Hamiltonian in the rest frame is described by
\bea
\mathcal{H}(x,t)=\lambda\omega_0\hat{a}^\dagger\hat{a}+\int_0^{+\infty}A(k,t)\cos[kx+\phi(k,\omega_d t)]dk.
\eea 
By transforming into the rotating frame with frequency $\Omega=\frac{\omega_d}{q}$ ($q\in \mathbb{Z}^+$) with free time-evolution operator $\hat{O}(t)\equiv e^{i\hat{a}^\dagger\hat{a}\frac{\omega_d}{q} t}$ and using 
$\hat{O}(t)\hat{x}\hat{O}^\dagger(t)=\hat{x}\cos (\Omega t)+\hat{p}\sin (\Omega t)$, we have the Hamiltonian in the rotating frame 
\bea\label{}
\hat{H}(t)&\equiv&\hat{O}(t)\mathcal{H}(t)\hat{O}^\dagger(t)-i\lambda O(t)\dot{O}^\dagger(t)\nl
&=& \mathcal{H}\Big[\hat{x}\cos \Big(\frac{\omega_d}{q} t\Big)+\hat{p}\sin \Big(\frac{\omega_d}{q} t\Big),t\Big]-\lambda\frac{\omega_d}{q}\hat{a}^\dagger\hat{a}\nl
&=&\lambda\Big(\omega_0-\frac{\omega_d}{q}\Big)\hat{a}^\dagger\hat{a}+\frac{1}{2}\int_0^{+\infty}A(k,t)\Big[e^{i\phi(k,\omega_d t)}e^{ik[\hat{x}\cos (\frac{\omega_d}{q} t)+\hat{p}\sin (\frac{\omega_d}{q} t)]}+h.c.\Big]dk.
\eea
In Flqoquet theory, the quasienergy operator of the time-periodic Hamiltonian $\hat{H}(t)$ is defined as $$\mathbb{\hat{H}}(t)\equiv \hat{H}(t)-i\lambda\partial/\partial t.$$ To calculate the eigenlevels and eigenstates of quasienergy operator $\mathbb{H}$,  we introduce the composite Hilbert space $\mathbb{F}\otimes\mathbb{T}$ that is a product of the Fock space   $\mathbb{F}=\{|m\rangle |m=0,1,\cdots \}$  and the temporal space $\mathbb{T}=\{|e^{iMt}\rangle | M=0,\pm 1, \pm 2, \cdots\}$. The matrix elements of quasienergy operator in this composite Hilbert are given by
\bea
&&\langle n, e^{iN\frac{\omega_d}{q}t}|\mathbb{\hat{H}}(t)| m, e^{iM\frac{\omega_d}{q}t}\rangle\nl
&=&\lambda\Big[\omega_0-(m-M)\frac{\omega_d}{q}\Big]\delta_{n,m}\delta_{N,M}\nl
&&+\frac{1}{2}\int_0^{+\infty}dk\langle e^{iN\frac{\omega_d}{q}t}|A(k,t)e^{i\phi(k,\omega_d t)}\langle n|e^{ik[\hat{x}\cos (\frac{\omega_d}{q} t)+\hat{p}\sin (\frac{\omega_d}{q} t)]}|m\rangle|e^{iM\frac{\omega_d}{q}t}\rangle\nl
&&+\frac{1}{2}\int_0^{+\infty}dk\langle e^{iN\frac{\omega_d}{q}t}|A(k,t)e^{-i\phi(k,\omega_d t)}\langle n|e^{-ik[\hat{x}\cos (\frac{\omega_d}{q} t)+\hat{p}\sin (\frac{\omega_d}{q} t)]}|m\rangle|e^{iM\frac{\omega_d}{q}t}\rangle\nl
&=&\lambda\Big[\omega_0-(m-M)\frac{\omega_d}{q}\Big]\delta_{n,m}\delta_{N,M}\nl
&&+\frac{1}{2}\int_0^{+\infty}dk\langle e^{i(N-n)\frac{\omega_d}{q}t}|A(k,t)e^{i\phi(k,\omega_d t)}|e^{i(M-m)\frac{\omega_d}{q}t}\rangle e^{-\frac{\lambda}{4}k^2}\Big(ik\sqrt{\frac{\lambda}{2}} \ \Big)^{m-n}\sqrt{\frac{n!}{m!}}L^{m-n}_n\Big(\frac{\lambda}{2}k^2\Big)\nl
&&+\frac{1}{2}\int_0^{+\infty}dk\langle e^{i(N-n)\frac{\omega_d}{q}t}|A(k,t)e^{-i\phi(k,\omega_d t)}|e^{i(M-m)\frac{\omega_d}{q}t}\rangle e^{-\frac{\lambda}{4}k^2}\Big(-ik\sqrt{\frac{\lambda}{2}} \ \Big)^{m-n}\sqrt{\frac{n!}{m!}}L^{m-n}_n\Big(\frac{\lambda}{2}k^2\Big),
\eea
where we have used the identity
\bea
\langle n|e^{ik[\hat{x}\cos (\frac{\omega_d}{q} t)+\hat{p}\sin (\frac{\omega_d}{q} t)]}|m\rangle&=&e^{-\frac{\lambda}{4}k^2-i(m-n)\frac{\omega_d }{q}t}\Big(ik\sqrt{\frac{\lambda}{2}} \ \Big)^{m-n}\sqrt{\frac{n!}{m!}}L^{m-n}_n\Big(\frac{\lambda}{2}k^2\Big)
\eea
and defined
\bea
\langle e^{i(N-n)\frac{\omega_d}{q}t}|A(k,t)e^{\pm i\phi(k,\omega_d t)}|e^{i(M-m)\frac{\omega_d}{q}t}\rangle&\equiv&\frac{\omega_d}{2\pi q}\int_0^{\frac{2\pi q}{\omega_d}}A(k,t)e^{\pm i\phi(k,\omega_d t)}\exp\Big[i(M-m-N+n)\frac{\omega_d}{q}t\Big]dt.
\eea
In the calculation of matrix elements, to avoid the divergence from $k^{m-n}$ when $k\rightarrow 0$ for $n>m$, one can use the identity
\bea
L^{n-m}_{m}(x)=\frac{n!}{m!}L^{m-n}_{n}(x)(-x)^{m-n}\ \ \ \mathrm{for}\ \ \  x>0.
\eea


\twocolumngrid

\bibliographystyle{unsrt}

\end{document}